\documentclass[doublecolumn, journal]{IEEEtran}

\usepackage{algorithm}
\usepackage{algpseudocode}
\usepackage{amsmath,amssymb,amscd}
\usepackage{graphicx,cite}
\usepackage{subfigure}

\begin{document}
\title{Multistatic Cloud Radar Systems: Joint Sensing and Communication Design}
\author{\IEEEauthorblockN{Seongah Jeong, Shahrouz Khalili, Osvaldo Simeone, Alexander Haimovich and Joonhyuk Kang}
\thanks{Seongah Jeong and Joonhyuk Kang are with the Department of Electrical Engineering, Korea Advanced Institute of Science and Technology (KAIST), Daejeon, South Korea.

Osvaldo Simeone and Alexander Haimovich are with the Center for Wireless Communications and Signal Processing Research (CWCSPR), ECE Department, New Jersey Institute of Technology (NJIT), Newark, NJ 07102, USA. 
}\\
}
\maketitle
\begin{abstract}
In a multistatic cloud radar system, receive sensors measure signals sent by a transmit element and reflected from a target and possibly clutter, in the presence of interference and noise. The receive sensors communicate over non-ideal backhaul links with a fusion center, or cloud processor, where the presence or absence of the target is determined. The backhaul architecture can be characterized either by an orthogonal-access channel or by a non-orthogonal multiple-access channel. Two backhaul transmission strategies are considered, namely compress-and-forward (CF), which is well suited for the orthogonal-access backhaul, and amplify-and-forward (AF), which leverages the superposition property of the non-orthogonal multiple-access channel. In this paper, the \emph{joint} optimization of the sensing and backhaul communication functions of the cloud radar system is studied. Specifically, the transmitted waveform is jointly optimized with backhaul quantization in the case of CF backhaul transmission and with the amplifying gains of the sensors for the AF backhaul strategy. In both cases, the information-theoretic criterion of the Bhattacharyya distance is adopted as a metric for the detection performance. Algorithmic solutions based on successive convex approximation are developed under different assumptions on the available channel state information (CSI). Numerical results demonstrate that the proposed schemes outperform conventional solutions that perform \emph{separate} optimizations of the waveform and backhaul operation, as well as the standard distributed detection approach.
\end{abstract}


\section{Introduction}
\begin{figure}[t]
\begin{center}
\includegraphics[width=7.5cm, height=4.5cm]{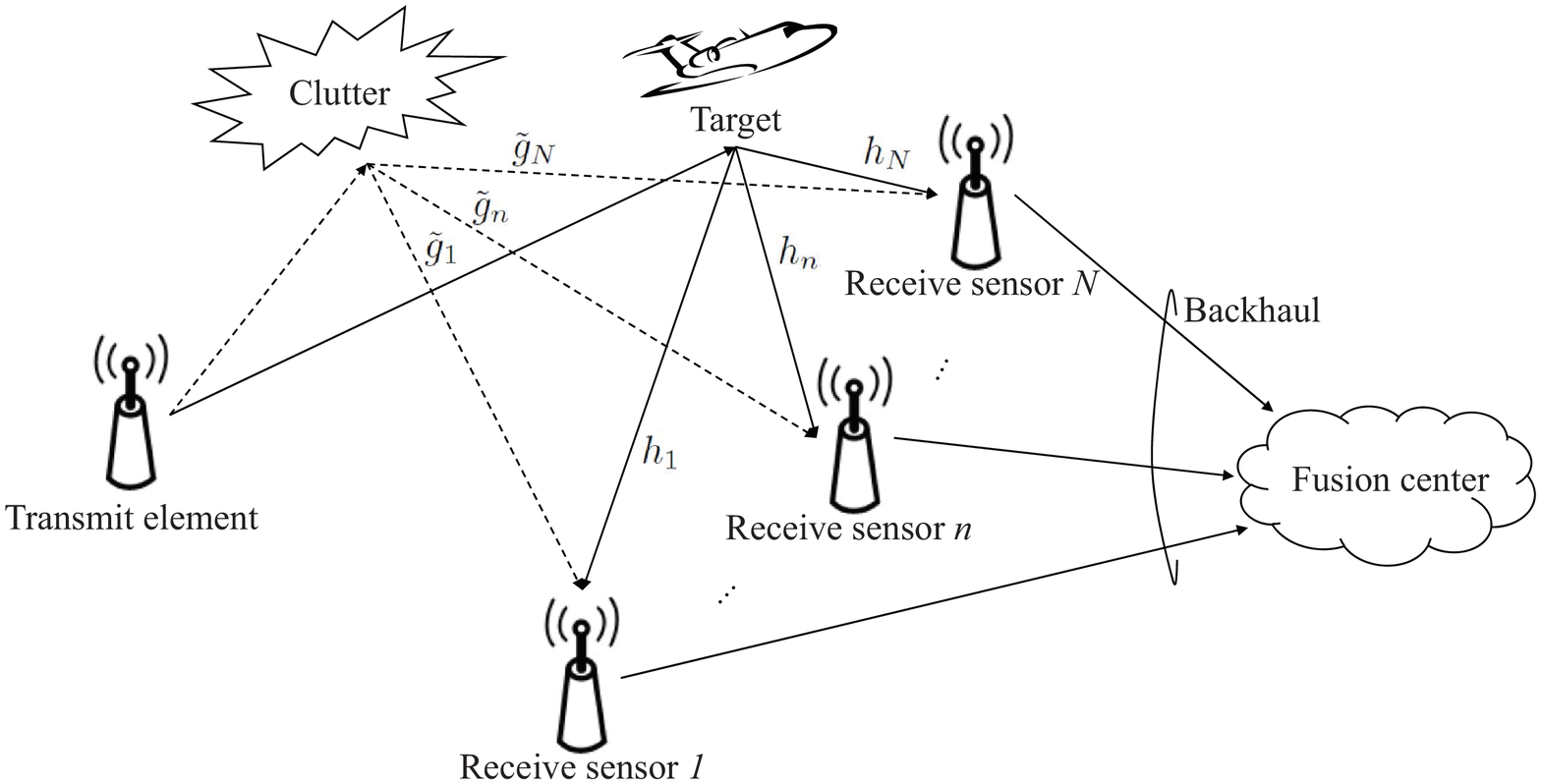}
\caption{Illustration of a multistatic cloud radar system, which consists of a transmit element, $N$ receive sensors, and a fusion center. All the nodes are configured with a single antenna. The receive sensors are connected to the fusion center via orthogonal-access or non-orthogonal multiple-access backhaul links.}\label{fig:sys}
\end{center}
\end{figure}

This paper\footnote{This work was partially presented at IEEE Radar Conference, Arlington, VA, May 2015 \cite{JSARadarCon15}.} addresses a distributed radar system that involves sensing and communication: a transmit element illuminates an area of interest, in which a target may be present, and the signals returned from the target are observed by sensors. The sensors have a minimal processing capabilities, but communicate over a backhaul network with a processing center, referred to henceforth as a \emph{fusion center}, where target detection takes place (see Fig. \ref{fig:sys}). Such architecture is different from a classical multistatic radar system in which each constituent radar performs the full array of radar functions, including target detection and tracking. The considered architecture is motivated by the proliferation of low-cost, mobile or fixed sensors in the ``Internet of Things,'' which are supported by global synchronization services such as the global positioning system (GPS), and are capable of communicating with a fusion center in the ``\emph{cloud}'' through a backhaul wireless or wired network. For example, the receive sensors could be mounted on light poles, trucks or unmanned aerial vehicles (UAV's) and could be connected to a wireless access point via Wi-Fi or dedicated mmWave links. As this architecture can be implemented by means of cloud computing technology, we refer to it as \textquotedblleft \emph{cloud radar}.\textquotedblright\ 

The main purpose of this work is to study the interaction between the sensing and backhaul communication functions in a cloud radar architecture, and to develop an understanding of the performance gain to be expected by means of a joint optimization of these two functions, namely of waveform design for sensing and of backhaul transmission.
\subsection{Background}\label{sec:background}
The separate design of radar waveforms, under the assumption of an ideal backhaul, has long been a problem of great interest \cite{Bernfeld12Book, Rihaczek67Book}. For monostatic radar systems, i.e., radars with single transmit and receive elements, optimal waveforms for detection in the Neyman-Pearson sense were studied in \cite{Kay07JSTSP}. In a multistatic radar system, where the signals received by a set of distributed sensors are processed jointly, the performance of the Neyman-Pearson optimal detector is in general too complex to be suitable as a design metric. As a result, various information-theoretic criteria such as the Bhattacharyya distance, the Kullback Leibler divergence, the J-divergence and the mutual information, which can be shown to provide various bounds to the probability of error (missed detection, false alarm and Bayesian risk), have been considered as alternative design metrics \cite{Kailath67TCOM, Kay09Aero,Stoica13TSP}.

Instead, the separate design of backhaul communication functions, for fixed radar waveforms was studied in \cite{Chen98,Barkat89Aero,Vis97ProcIEEE,Blum97ProcIEEE} under a compress-and-forward (CF) strategy, for which backhaul quantization was optimized, and in \cite{Ali14TWC, Ali14sensor} under an amplify-and-forward (AF) scheme, for which the power allocation at the sensors was investigated using the minimum mean square error (MMSE) as the performance criterion. 
\subsection{Main Contributions}\label{sec:main}
\textit{Unlike prior work, in this paper, we tackle the problem of \textit{jointly} designing the waveform, or code vector, and the transmission of the receive sensors over the backhaul.} This approach is motivated by the strong interplay between waveform and backhaul transmission designs. For instance, waveform design may allocate more power at frequencies that are less affected on average by clutter and interference, while the backhaul transmission strategies are adapted accordingly to devote most backhaul resources, namely capacity or power, to the transmission of such frequencies to the fusion center. 

Two basic types of backhaul links between the radar receive sensors and the fusion center are considered, namely orthogonal and non-orthogonal access backhaul. In the former, no interference exists between the sensors, as in a wired backhaul, while in the latter, the backhaul forms a multiple-access channel, where channels are subject to mutual interference, as in a wireless backhaul. Furthermore, two standard backhaul transmission schemes are investigated, namely CF and AF. As in the Cloud Radio Access Network (C-RAN) architecture in communication \cite{ChinaMobile}, CF is particularly well suited to an orthogonal backhaul architecture: each sensor satisfies the backhaul capacity constraint quantizing the received baseband signals prior to transmission to the fusion center. AF, instead, is better matched to a non-orthogonal multiple-access backhaul: each receive sensor amplifies and forwards the received signal to the fusion center so that the signals transmitted by the receive sensors are superimposed at the fusion center (see, e.g., \cite{Ali14TWC, Ali14sensor}).

Our specific contributions are as follows:\\
$\bullet$ CF: The joint optimization of the waveform and the quantization strategy is investigated for CF, with a focus on orthogonal-access backhaul. To reflect practical constraints, only stochastic channel state information (CSI) is assumed on the channel gains between target or clutter and the receive sensors. For an optimization objective, we adopt the information-theoretic criterion of the Bhattacharyya distance in order to account for the detection performance \cite{Kailath67TCOM, Kay09Aero, Stoica13TSP}.

$\bullet$ AF: The joint optimization of the waveform and the amplifying gains of the receive sensors is studied for AF, by concentrating on non-orthogonal multiple-access backhaul. We adopt the performance criterion and main assumptions of CF. Furthermore, we consider both instantaneous and stochastic CSI on the receive sensors-to-fusion center channels. 

Throughout, we assume tractable and well accepted models in order to gain insight into the problem at hand. With this insight gained, subsequent work may explore more detailed configurations.

The rest of the paper is organized as follows. In Section \ref{sec:sm}, we present the signal model and cover the two types of backhaul links, namely orthogonal-access and non-orthogonal multiple-access backhaul. In Section \ref{sec:CFopt}, after describing the CF backhaul transmission strategies and reviewing the optimal detectors, we present the optimization of the multistatic cloud radar system with CF. In Section \ref{sec:AFopt}, we focus on the AF backhaul transmission, and optimize the system with both instantaneous and stochastic CSI under AF. Numerical results are provided in Section \ref{sec:num}, and, finally, conclusions are drawn in Section \ref{sec:con}. 
\section{System Model}\label{sec:sm}
Consider a multistatic cloud radar system consisting of a transmit element, $N$ receive sensors, and a fusion center, or cloud processor, as illustrated in Fig. \ref{fig:sys}. The receive sensors communicate with the fusion center over an orthogonal-access backhaul or a non-orthogonal multiple-access backhaul. All the nodes are equipped with a single antenna, and the set of receive sensors is denoted $\mathcal{N}=\{1,\dots ,N\}$.

The system aims to detect the presence of a single stationary target in a clutter field. To this end, each sensor receives a noisy version of the signal transmitted by the transmit element and reflected from the surveillance area, which is conveyed to the fusion center on the backhaul channels after either quantizing or amplifying the received signals as discussed below. It is assumed that perfect timing information is available at the fusion center, such that samples of the received signal may be associated with specific locations in some coordinate system. For such a location, and based on all the signals forwarded from the different receive sensors, the fusion center makes a decision about the presence of the target (see, e.g.\cite{Stoica13TSP, Chen98,Barkat89Aero,Vis97ProcIEEE,Blum97ProcIEEE,Ali14TWC, Ali14sensor}). Note that, as argued in \cite{Kay09Aero}, the assumption of stationary target and scatterers can be regarded as a worst-case scenario for more general set-ups with non-zero Doppler. 

We consider a pulse compression radar in which the transmitted signal given in baseband form is 
\begin{equation}\label{eq:txsig}
s(t)=\sum_{k=1}^{K}x_k\phi(t-(k-1)T_c),
\end{equation}
where $\phi(t)$ is, for example, a square root Nyquist with chip rate $1/T_c$, so that $\{\phi(t-(k-1)T_c)\}_{k=1}^K$ are orthonormal; and $\{x_k\}_{k=1}^K$ is a sequence of (deterministic) complex coefficients that modulate the waveform. The vector $\pmb{x}=[x_1\,\, \cdots \,\,x_K]^T$ is referred to as \textit{waveform} or \textit{code vector}, on which we impose the transmit power constraint $\pmb{x}^H\pmb{x} \le P_T$. The design of the waveform $\pmb{x}$ determines both target and clutter response, and thus has a key role in the performance of the radar system.

The baseband signal received at the sensor $n \in \mathcal{N}$, which is backscattered by a stationary target, can be expressed as
\vspace{-0.2cm}
\begin{equation}\label{eq:rxsig}
r_n(t)=h_ns(t-\tau_n)+c_n(t)+w_n(t),
\end{equation}
where $h_n$ is the random complex amplitude of the target return, which includes the effects of the channel and follows a Swerling I target-type model having a Rayleigh envelope, i.e., $h_n \sim \mathcal{CN}(0, \sigma_{t,n}^2)$; $c_n(t)$ represents the clutter component; $w_n(t)$ is a Gaussian random process representing the signal-independent interference, which aggregates the contributions of thermal noise, interference and jamming and is assumed to be correlated over time, as detailed below; and $\tau_n$ is the propagation delay for the path from the transmit element to the target and thereafter to the sensor $n$, which is assumed to satisfy the condition $\tau_n \ge KT_c$ in order for the target to be detectable. The clutter component $c_n(t)$ consists of signal echoes generated by stationary point scatterers, whose echoes have independent return amplitudes and arrival times. Accordingly, the clutter component $c_n(t)$ is expressed as
\vspace{-0.2cm}
\begin{equation}\label{eq:clutter}
c_n(t)=\sum_{v=1}^{N_c}g_{n,v}s(t-\tau_{n,v}),
\end{equation}
where $N_c$ is the number of point scatters; $g_{n,v}$ is the amplitude of the return from scatterer $v$; and $\tau_{n,v}$ is the propagation delay for the path from the transmit element to the scatterer $v$ and to the sensor $n$, which satisfies the condition $\tau_{n,v} \le KT_c$. 

After matched filtering of the received signal (\ref{eq:rxsig}) with the impulse response $\phi^*(-t)$, and after range-gating by sampling the output of the matched filter at the chip rate, the discrete-time signal at receive sensor $n$ for $n \in \mathcal{N}$ can be written as 
\begin{equation}\label{eq:rxdiscrete}
r_{n,k}=h_nx_k+\tilde{g}_nx_k+w_{n,k},
\end{equation} 
where $r_{n,k}$ is the output of the matched filter at the receive sensor $n$ sampled at time $t=(k-1)T_c+\tau_n$; the term $\tilde{g}_n=\sum_{v=1}^{N_c}g_{n,v}\Psi(\tau_{n}-\tau_{n,v})$ with  $\Psi(t)\triangleq\int_{-\infty}^{\infty}\phi(\tau-t)\phi^*(\tau)d\tau$ being the auto-correlation function of $\phi(t)$, represents the contribution of clutter scatterers, which can be modeled, invoking the central limit theorem, as a zero mean Gaussian random variable with a given variance $\sigma_{c,n}^2$ (see \cite[Appendix A]{Stoica13TSP}); and $w_{n,k}$ is the $k$th sample of $w_n(t)$ after matched filtering at the sensor $n$. 

In vector notation, we can write (\ref{eq:rxdiscrete}) as
\vspace{-0.2cm}
\begin{equation}\label{eq:rxvector}
\pmb{r}_n=\pmb{s}_n+\pmb{c}_n+\pmb{w}_n,
\end{equation}  
where we defined $\pmb{r}_n\triangleq[r_{n,1}\,\,\cdots\,\,r_{n,K}]^T$, $\pmb{s}_n\triangleq h_n\pmb{x}$ and $\pmb{c}_n\triangleq\tilde{g}_n\pmb{x}$; and the noise vector $\pmb{w}_n\triangleq[w_{n,1}\,\,\cdots\,\,w_{n,K}]^T$ follows a zero-mean Gaussian distribution with temporal correlation $\pmb{\Omega}_{w,n}$, i.e., $\pmb{w}_n \sim \mathcal{CN}(\pmb{0}, \pmb{\Omega}_{w,n})$. The variables $h_n$, $\tilde{g}_n$ and $\pmb{w}_n$ for all $n \in \mathcal{N}$, are assumed to be independent for different values of $n$ under the assumption that the receive sensors are sufficiently separated \cite{Kay09Aero}. Moreover, their second-order statistics $\sigma_{t,n}^{2}$, $\sigma_{c,n}^{2}$ and $\pmb{\Omega}_{w,n}$ are assumed to be known to the fusion center, for all $n \in \mathcal{N}$, e.g., from prior measurements or prior information \cite{Gini08,Gue10Radar}.

To summarize, the signal received at sensor $n$ can be written as
\vspace{-0.2cm}
\begin{subequations}\label{eq:HypoRA}
\begin{eqnarray}
&&\mathcal{H}_0 : \pmb{r}_n=\pmb{c}_n+\pmb{w}_n, \\
&&\mathcal{H}_1 : \pmb{r}_n=\pmb{s}_n+\pmb{c}_n+\pmb{w}_n, \hspace{0.2cm} n \in \mathcal{N},
\end{eqnarray}         
\end{subequations}
where $\mathcal{H}_0$ and $\mathcal{H}_1$ represent the hypotheses under which the target is absent or present, respectively. 

In the rest of this section, we detail the assumed model for both orthogonal-access and non-orthogonal multiple-access backhaul. 

{\bf{Orthogonal-access Backhaul:}} For the orthogonal-access backhaul case, each receive sensor $n$ is connected to the fusion center via an orthogonal link of limited capacity $C_n$ bits per received sample. The capacity $C_n$ is assumed to be known to the fusion center for all $n \in \mathcal{N}$ and to change sufficiently slowly so as to enable the adaptation of the waveform and of the transmission strategy of the sensors to the values of the capacities $C_n$ for all $n \in \mathcal{N}$. 

{\bf{Non-orthogonal Multiple-access Backhaul:}} For the non-orthogonal multiple-access backhaul, the signal received at the fusion center is the superposition of the signals sent by all receive sensors, where channels are subject to mutual interference. Accordingly, the received signal at the fusion center $\tilde{\pmb{r}}=[\tilde{r}_1 \,\,\cdots\,\,\tilde{r}_K]^T$ is given by 
\begin{equation}\label{eq:multiFC}
\tilde{\pmb{r}}=\sum_{n=1}^Nf_n\pmb{t}_n+\pmb{z},
\end{equation}
where  $\pmb{t}_n=[t_{n,1} \,\,\cdots\,\,t_{n,K}]^T$ is the signal sent by the receive sensor $n$ on the backhaul to the fusion center; $f_n$ is the complex-valued channel gain between the receive sensor $n$ and the fusion center; and $\pmb{z}=[z_1\,\,\cdots\,\,z_K]^T$ is the noise vector having a zero-mean Gaussian distribution with correlation matrix $\pmb{\Omega}_{z}$, i.e., $\pmb{z} \sim\mathcal{CN}(\pmb{0},\pmb{\Omega}_z)$. Based on prior information or measurements, the second-order statistics of the channel gains between the target and the receive sensors, and of the noise terms, namely $\sigma_{t,n}^2$, $\sigma_{c,n}^2$, $\pmb{\Omega}_{w,n}$ and $\pmb{\Omega}_z$, are assumed to be known to the fusion center for all $n \in \mathcal{N}$. The channel between receive sensors and fusion center $\pmb{f}=[f_1 \,\,\cdots\,\,f_N]^T$ are also assumed to be known at the fusion center, via training and channel estimation.
\section{CF Backhaul Transmission}\label{sec:CFopt}
In this section, we consider orthogonal-access backhaul and CF transmission. With CF, each receive sensor quantizes the received vector $\pmb{r}_{n}$ in (\ref{eq:HypoRA}), and sends a quantized version of $\pmb{r}_n$ to the fusion center. Note that, since the receive sensor does not know whether the target is present or not, the quantizer cannot depend on the correct hypothesis $\mathcal{H}_{0}$ or $\mathcal{H}_{1}$. In order to facilitate analysis and design, we follow the standard random coding approach of rate-distortion theory of modeling the effect of quantization by means of an additive quantization noise (see, e.g., \cite{Gersho92, Cov06}) as in 
\vspace{-0.2cm}
\begin{equation}\label{eq:CF}
\tilde{\pmb{r}}_n=\pmb{r}_n+\pmb{q}_n,
\end{equation}
where $\tilde{\pmb{r}}_n=[\tilde{r}_{n,1}\,\,\cdots\,\,\tilde{r}_{n,K}]^T$ is the quantized signal vector of $\pmb{r}_n$; and $\pmb{q}_{n}\sim\mathcal{CN}(\pmb{0},\pmb{\Omega}_{q,n})$ is the quantization error vector, which is characterized by a covariance matrix $\pmb{\Omega}_{q,n}$. Based on random coding arguments, while (\ref{eq:CF}) holds on an average over randomly generated  quantization codebooks, the results derived in this paper can be obtained by means of some (deterministic) high-dimensional vector quantizer (see, e.g., \cite{GamalBook}). For instance, as discussed in \cite{Zamir96TIT}, a Gaussian quantization noise $\pmb{q}_n$ with any covariance $\pmb{\Omega}_{q,n}$ can be realized in practice via a linear transform, obtained from the eigenvectors of $\pmb{\Omega}_{q,n}$, followed by a multi-dimensional dithered lattice quantizer such as Trellis Coded Quantization (TCQ) \cite{Marce90TCOM}. 

Based on (\ref{eq:CF}), the signal received at the fusion center from receive sensor $n$ is given as
\vspace{-0.2cm}
\begin{equation}
\begin{split} & \mathcal{H}_{0}:~\tilde{\pmb{r}}_{n}=\pmb{c}_{n}+\pmb{w}_{n}+\pmb{q}_{n},\\
 & \mathcal{H}_{1}:~\tilde{\pmb{r}}_{n}=\pmb{s}_{n}+\pmb{c}_{n}+\pmb{w}_{n}+\pmb{q}_{n}.
\end{split}\label{eq:ortho_FCNW}
\end{equation}
As further elaborated in the following, the covariance matrix $\pmb{\Omega}_{q,n}$ determines the bit rate required for backhaul communication between the receive sensor $n$ and the fusion center \cite{Gersho92,Cov06} and is subject to design. 

To set the model (\ref{eq:ortho_FCNW}) in a more convenient form, the signal received at the fusion center is whitened with respect to the overall additive noise $\pmb{c}_{n}+\pmb{w}_{n}+\pmb{q}_{n}$, and the returns from all sensors are collected, leading to the model
\vspace{-0.2cm}
\begin{equation}
\begin{split} & \mathcal{H}_{0}:~\pmb{y}\sim\mathcal{CN}(\pmb{0},\pmb{I}),\\
 & \mathcal{H}_{1}:~\pmb{y}\sim\mathcal{CN}(\pmb{0},\pmb{DSD}+\pmb{I}),
\end{split}\label{eq:ortho_FCW}
\end{equation}
where $\pmb{y}=[\pmb{y}_{1}^{T}\,\,\cdots\,\,\pmb{y}_{N}^{T}]^{T},$ $\pmb{y}_{n}=\pmb{D}_{n}\tilde{\pmb{r}}_{n},$ $\pmb{D}_{n}\pmb{\ }$ is the whitening matrix associated with the receive sensor $n$ and is given by $\pmb{D}_{n}=(\sigma_{c,n}^{2}\pmb{x}\pmb{x}^{H}+\pmb{\Omega}_{w,n}+\pmb{\Omega}_{q,n})^{-1/2},$ $\pmb{D}$ is the block diagonal matrix $\pmb{D}=\text{diag}\{\pmb{D}_{1},...,\pmb{D}_{N}\}$, and $\pmb{S}$ is the block diagonal matrix $\pmb{S}=\text{diag}\{\sigma_{t,1}^{2}\pmb{x}\pmb{x}^{H},...,\sigma_{t,N}^{2}\pmb{x}\pmb{x}^{H}\}$. The detection problem formulated in (\ref{eq:ortho_FCW}) has the standard Neyman-Pearson solution given by the test \vspace{-0.2cm}
\begin{align}\label{eq:ortho_test}
        							&\mathcal{H}_1 \nonumber\\[-4pt]
\pmb{y}^H\pmb{T}\pmb{y}        &\gtreqless   \nu, \\[-4pt]
        							&\mathcal{H}_0 \nonumber
\end{align} 
where we have defined $\pmb{T}=\pmb{D}\pmb{S}\pmb{D}(\pmb{D}\pmb{S}\pmb{D}+\pmb{I})^{-1}$, and the threshold $\nu$ is set based on the tolerated false alarm probability \cite{Kay93Book}.

In the rest of this section, we aim to find the optimum code vector $\pmb{x}$ and quantization error covariance matrices $\pmb{\Omega}_{q,n}$ in (\ref{eq:ortho_FCNW}), for given backhaul capacity constraints $C_n$, for all $n \in \mathcal{N}$. Before we proceed, for reference, we first discuss the standard distributed detection approach that combines hard local decisions at the receive sensors and a majority-rule detection at the fusion center (see, e.g., \cite{Varshney97, Aky11physical}). 
\subsection{Distributed Detection}\label{sec:dist}
Here, we describe the standard distributed detection approach applied to multistatic radar system (see, e.g., \cite{Varshney97, Aky11physical}). With this approach, each receive sensor $n$ makes its own decision based on the likelihood test given by $\pmb{y}_n^H\pmb{T}_n\pmb{y}_n \overset {\mathcal{H}_1}{\underset {\mathcal{H}_0}{\gtreqless}} \gamma_n$, where $\gamma_n$ is the threshold for receive sensor $n$, which is calculated based on the tolerated false alarm probability \cite{Kay93Book}, and we have defined $\pmb{T}_n=\pmb{D}_n\pmb{S}_n\pmb{D}_n(\pmb{D}_n\pmb{S}_n\pmb{D}_n+\pmb{I})^{-1}$ with $\pmb{y}_n=\pmb{D}_n\pmb{r}_n$, $\pmb{D}_n=(\sigma_{c,n}^2\pmb{x}\pmb{x}^H+\pmb{\Omega}_{w,n})^{-1/2}$ and $\pmb{S}_n=\sigma_{t,n}^2\pmb{x}\pmb{x}^H$, for all $n \in \mathcal{N}$. The receive sensors transmit the obtained one-bit hard decision to the fusion center. Note that this scheme is feasible as long as the backhaul capacity available for each receive sensor-to-fusion center channel is larger than or equal to $1/K$ bits/sample, i.e., $C_n \ge 1/K$, for $n \in \mathcal{N}$. The fusion center decides on the target's presence based on the majority rule: if the number of receive sensors $k$ that decide for $\mathcal{H}_0$ satisfies $k \ge N/2$, the fusion center chooses $\mathcal{H}_0$, and vice versa if $k \le N/2$. 
\subsection{Performance Metrics and Constraints}\label{sec:cloud}
To start the analysis of the cloud radar system, we discuss the criterion that is adopted to account for the detection performance, namely the Bhattacharyya distance and the approach used to model the effect of the quantizers at the receive sensors. 

{\bf{Bhattacharyya Distance:}} For two zero-mean Gaussian distributions with covariance matrix of $\pmb{\Sigma}_{1}$ and $\pmb{\Sigma}_{2},$ the Bhattacharyya
distance $\mathcal{B}$ is given by \cite{Kailath67TCOM}
\vspace{-0.2cm}
\begin{equation}\label{eq:Bdist}
\mathcal{B}=\log\left(\frac{\left|0.5(\boldsymbol{\Sigma}_{1}+\boldsymbol{\Sigma}_{2})\right|}{\sqrt{|\boldsymbol{\Sigma}_{1}||\boldsymbol{\Sigma}_{2}|}}\right).
\end{equation}
Therefore, for the signal model (\ref{eq:ortho_FCNW}), the Bhattacharyya distance between the distributions under the two hypotheses can be calculated as
\vspace{-0.2cm}
\begin{eqnarray}\label{eq:ortho_Bdist}
&&\hspace{-0.9cm}\mathcal{B}(\pmb{x}, \pmb{\Omega}_q)=\log\left(\frac{\left|\pmb{I}+0.5\pmb{D}\pmb{S}\pmb{D}\right|}{\sqrt{\left|\pmb{I}+\pmb{D}\pmb{S}\pmb{D}\right|}}\right)\nonumber\\
&&\hspace{0.35cm}=\sum_{n=1}^{N}\mathcal{B}_{n}(\pmb{x},\pmb{\Omega}_{q,n})\nonumber\\
&&\hspace{0.35cm}=\sum_{n=1}^{N}\log\left(\frac{1+0.5\lambda_{n}}{\sqrt{1+\lambda_{n}}}\right),
\end{eqnarray}
where we have made explicit the dependence on $\pmb{x}$ and $\pmb{\Omega}_{q,n}$; $\pmb{\Omega}_q$ collects all the covariance matrices of quantization noise and is given as $\pmb{\Omega}_q=\{\pmb{\Omega}_{q,n}\}_{n \in \mathcal{N}}$; and we have defined
\vspace{-0.2cm}
\begin{equation}\label{eq:ortho_lam}
\lambda_{n}=\sigma_{t,n}^{2}\pmb{x}^{H}\left(\sigma_{c,n}^{2}\pmb{x}\pmb{x}^{H}+\pmb{\Omega}_{w,n}+\pmb{\Omega}_{q,n}\right)^{-1}\pmb{x}.
\end{equation}
We observe that (\ref{eq:ortho_Bdist}) is valid under the assumption that the effect of the quantizers can be well approximated by additive Gaussian noise as per (\ref{eq:ortho_FCNW}). This is discussed next. 

{\bf{Quantization:}} From rate-distortion theory, a vector quantizer exists that is able to realize the additive quantization noise model (\ref{eq:CF}), when operating over a sufficiently large number of measurement vectors (\ref{eq:HypoRA}), as long as the capacity $C_n$ is no smaller than the mutual information $I(\pmb{r}_{n};\tilde{\pmb{r}}_{n})/K$ \cite{GamalBook}. For example, a dithered lattice vector quantizer achieves this result \cite{Zamir96TIT}. These considerations motivate the selection of the mutual information $I(\pmb{r}_{n};\tilde{\pmb{r}}_{n})$ as a measure of the backhaul rate required for the transmission to the fusion center.
 
While the mutual information $I(\pmb{r}_{n};\tilde{\pmb{r}}_{n})$ depends on the actual hypothesis $\mathcal{H}_{0}$ or $\mathcal{H}_{1}$, it is easy to see that $I(\pmb{r}_{n};\tilde{\pmb{r}}_{n})$ is larger under hypothesis $\mathcal{H}_{1}$. Based on this, the mutual information $I(\pmb{r}_{n};\tilde{\pmb{r}}_{n})$ evaluated under $\mathcal{H}_{1}$ is adopted here as the measure of the bit rate required between receive sensor $n$ and the fusion center. This can be easily calculated as $I(\pmb{r}_{n};\tilde{\pmb{r}}_{n})=\mathcal{I}_{n}(\pmb{x},\pmb{\Omega}_{q,n})$ by using the expression of the mutual information for multivariate Gaussian distribution (see, e.g., \cite{Cov06}) with
\vspace{-0.2cm}
\begin{eqnarray}\label{eq:ortho_rate}
&&\hspace{-1.2cm}\mathcal{I}_{n}(\pmb{x},\pmb{\Omega}_{q,n}) =\log\left|\pmb{I}+(\pmb{\Omega}_{q,n})^{-1}\pmb{\Omega}_{w,n}\right|\nonumber\\
&&\hspace{-1.2cm}+\log\left(1+(\sigma_{t,n}^{2}+\sigma_{c,n}^{2})\pmb{x}^{H}(\pmb{\Omega}_{w,n}+\pmb{\Omega}_{q,n})^{-1}\pmb{x}\right),
\end{eqnarray}
where again we have made explicit the dependence of mutual information on $\pmb{x}$ and $\pmb{\Omega}_{q,n}$. 

In the following, we formulate and solve the problem of jointly optimizing the Bhattacharyya distance criterion over the waveform $\pmb{x}$ at the transmit element and over the covariance matrices $\pmb{\Omega}_q$ of the quantizers at the receive sensors in Section \ref{sec:ortho_prob} and in Section \ref{sec:ortho_algo}, respectively.    
\subsection{Problem Formulation}\label{sec:ortho_prob}
The problem of maximizing the Bhattacharyya distance in (\ref{eq:ortho_Bdist}) over the waveform $\pmb{x}$ and the covariance matrices $\pmb{\Omega}_q$ under the backhaul capacity constraints is stated as 
\vspace{-0.2cm}
\begin{subequations}\label{eq:ortho_opt}
\begin{eqnarray}
&&\hspace{-1.2cm}{\mathop {\text{minimize} }\limits_{\pmb{x},\pmb{\Omega}_q}} \hspace{0.15cm}{\bar{\mathcal{B}}(\pmb{x}, \pmb{\Omega}_q)=\sum_{n=1}^N\bar{\mathcal{B}}_n(\pmb{x}, \pmb{\Omega}_{q_n})}\label{eq:ortho_optObj}\\
&& \hspace{-0.8cm}{\rm{s.t.}}\hspace{0.5cm} {\mathcal{I}_{n}(\pmb{x},\pmb{\Omega}_{q,n})\le KC_n=\bar{C}_n, \hspace{0.2cm} n \in \mathcal{N},}\label{eq:ortho_optRate}\\
&& \hspace{0.15cm} {\pmb{x}^H\pmb{x} \le P_T,} \\
&& \hspace{0.15cm} {\pmb{\Omega}_{q,n} \succeq 0, \hspace{0.2cm} n \in \mathcal{N},}
\end{eqnarray}
\end{subequations}
where we have formulated the problem as the minimization of the negative distance $\bar{\mathcal{B}}(\pmb{x},\pmb{\Omega}_q)=\sum_{n=1}^{N}$ $\bar{\mathcal{B}}_{n}(\pmb{x},\pmb{\Omega}_{q,n})$, with $\bar{\mathcal{B}}(\pmb{x},\pmb{\Omega}_q)=-\mathcal{B}(\pmb{x},\pmb{\Omega}_q)$ and $\bar{\mathcal{B}}_{n}(\pmb{x},\pmb{\Omega}_{q,n})=-\mathcal{B}_{n}(\pmb{x},\pmb{\Omega}_{q,n})$, following the standard convention in \cite{Boyd04Book}. The power of the waveform $\pmb{x}$ is constrained not to exceed a prescribed value of transmit power $P_T$. We observe that the constraint (\ref{eq:ortho_optRate}) ensures that the transmission rate with $K$ chips between each receive sensor and the fusion center is smaller than $\bar{C}_n$, according to the adopted information-theoretic metric. Note also that the problem (\ref{eq:ortho_opt}) is not a convex program, since the objective function (\ref{eq:ortho_optObj}) and the constraints (\ref{eq:ortho_optRate}) are not convex. 
\subsection{Proposed Algorithm}\label{sec:ortho_algo}
Since both functions $\bar{\mathcal{B}}_n(\pmb{x}, \pmb{\Omega}_{q_n})$ and $\mathcal{I}_n(\pmb{x}, \pmb{\Omega}_{q_n})$ in (\ref{eq:ortho_opt}) are non-convex in $\pmb{x}$ and $\pmb{\Omega}_{q,n}$, the optimization problem (\ref{eq:ortho_opt}) is not convex, and hence it is difficult to solve. To obtain a locally optimal solution, we approach the joint optimization of $\pmb{x}$ and $\pmb{\Omega}_q$ in (\ref{eq:ortho_opt}) via successive convex approximations. Specifically, in an outer loop,  Block Coordinate Descent (BCD) is applied to update $\pmb{x}$ and $\pmb{\Omega}_q$ one at a time, while an inner loop implemented via Majorization-Minimization (MM) solves the optimization of $\pmb{x}$ and $\pmb{\Omega}_q$ separately. This approach was first introduced in \cite{Osvaldo14TSPLett} for a sum-capacity backhaul constraint. By the properties of MM (see, e.g., \cite{Hunter04Amer, Luo13SIAM}), the algorithm provides a sequence of feasible solutions with non-increasing cost function, which guarantees convergence of the cost function. Note that, due to the non-convexity of the problem, no claim of convergence to a local or global optimum is made here.  

At the $i$th iteration of the outer loop, the optimum waveform $\pmb{x}^{(i)}$ is obtained by solving (\ref{eq:ortho_opt}) for matrices $\pmb{\Omega}_q=\pmb{\Omega}_q^{(i-1)}$ obtained at the previous iteration; subsequently, the matrices $\pmb{\Omega}_q^{(i)}$ are calculated by solving (\ref{eq:ortho_opt}) with $\pmb{x}=\pmb{x}^{(i)}$. These two separate optimizations are carried out by the MM method, which, as described in Appendix A, requires the solution of a quadratically constrained quadratic programs (QCQP). The proposed algorithm coupling BCD and MM to solve problem (\ref{eq:ortho_opt}), is summarized in Table Algorithm \ref{al1}. In Algorithm \ref{al1}, we use the superscript $i$ to identify the iterations of the outer loop, and the superscript $j$ as the index of the inner iteration of the MM method (e.g., $\pmb{x}^{(i,j)}$ indicates the waveform optimized at the $j$th iteration of the inner loop of the MM method and the $i$th iteration of the outer loop). In Appendix A, we present the MM steps and the overall proposed algorithm in detail. 

The complexity of Algorithm \ref{al1} by using standard convex optimization tools is polynomial in $K$ and $N$ since, at each outer iteration, MM requires to solve the problems (\ref{eq:ortho_opt_code}) and (\ref{eq:ortho_opt_quant}), whose sizes of the optimization domains are $K$ and $NK^2$, and numbers of constraints are $N+1$ and $2N$, respectively \cite{Luo10SPM, Boyd04Book}.

\renewcommand{\baselinestretch}{0.8}
\begin{algorithm} [h!]
\begin{algorithmic}
\caption{Joint optimization of waveform and quantization noise covariances (\ref{eq:ortho_opt})} \label{al1}
\State {\textbf{Initialization (outer loop)}}: Initialize $\pmb{x}^{(0)} \in C^{K \times 1}$, $\pmb{\Omega}_q^{(0)} \succeq 0$ and set $i=0$.
\State {\textbf{Repeat (BCD method)}}
\State \indent $i \gets i+1$
\State \indent {\textbf{Initialization (inner loop)}}: Initialize $\pmb{x}^{(i,0)}=$ 
\State \indent $\pmb{x}^{(i-1)}$ and set $j=0$.
\State \indent {\textbf{Repeat (MM method for $\pmb{x}^{(i)}$)}}
\State \indent \indent $j \gets j+1$
\State \indent \indent Find $\pmb{x}^{(i,j)}$ by solving the problem (\ref{eq:ortho_opt_code}) with 
\State \indent \indent $\pmb{\Omega}_q=\pmb{\Omega}_q^{(i-1)}$.
\State \indent {\textbf{Until}} a convergence criterion is satisfied.
\State \indent {\textbf{Update}} $\pmb{x}^{(i)} \gets \pmb{x}^{(i,j)}$
\State \indent {\textbf{Initialization (inner loop)}}: Initialize $\pmb{\Omega}_q^{(i,0)}=$ 
\State \indent $\pmb{\Omega}_q^{(i-1)}$ and set $j=0$. 
\State \indent {\textbf{Repeat (MM method for $\pmb{\Omega}_q^{(i)}$)}}
\State \indent \indent $j \gets j+1$
\State \indent \indent Find $\pmb{\Omega}_q^{(i,j)}$ by solving the problem (\ref{eq:ortho_opt_quant}) with 
\State \indent \indent $\pmb{x}=\pmb{x}^{(i)}$.
\State \indent {\textbf{Until}} a convergence criterion is satisfied.
\State \indent {\textbf{Update}} $\pmb{\Omega}_q^{(i)} \gets \pmb{\Omega}_q^{(i,j)}$
\State {\textbf{Until}} a convergence criterion is satisfied. 
\State {\textbf{Solution}}: $\pmb{x} \gets \pmb{x}^{(i)}$ and  $\pmb{\Omega}_q \gets \pmb{\Omega}_q^{(i)}$
\end{algorithmic}
\end{algorithm}
\renewcommand{\baselinestretch}{1}
\section{AF Backhaul Transmission}\label{sec:AFopt}
In this section, we consider AF transmission on a non-orthogonal multiple-access backhaul. With AF, sensor $n \in \mathcal{N}$ amplifies the received signal $\pmb{r}_n$ in (\ref{eq:HypoRA}) and then forwards the amplified signal $\pmb{t}_n=\alpha_n\pmb{r}_n$ to the fusion center, where $\alpha_n$ is is the amplification coefficient at the receive sensor $n$. From (\ref{eq:multiFC}), the fusion center is faced with the following detection hypothesis problem 
\vspace{-0.2cm}
\begin{eqnarray}\label{eq:multi_FCNW}
&&\hspace{-0.8cm}\mathcal{H}_{0}:~\tilde{\pmb{r}}=\sum_{n=1}^Nf_n\pmb{t}_n+\pmb{z}=\sum_{n=1}^Nf_n\alpha_n\left(\pmb{c}_{n}+\pmb{w}_{n}\right)+\pmb{z},\nonumber\\
&&\hspace{-0.8cm}\mathcal{H}_{1}:~\tilde{\pmb{r}}=\sum_{n=1}^Nf_n\pmb{t}_n+\pmb{z}\sum_{n=1}^Nf_n\alpha_n\left(\pmb{s}_{n}+\pmb{c}_{n}+\pmb{w}_{n}\right)+\pmb{z}.\nonumber\\
\end{eqnarray}
The variables $h_n$, $\tilde{g}_n$, $\pmb{w}_n$, $f_n$ and $\pmb{z}$, for all $n \in \mathcal{N}$, are assumed to be mutually independent. Since only the second-order statistics of the channel gains $h_n$, $n \in \mathcal{N}$, are known to the receive sensors and the fusion center, no coherent gains may be achieved by optimizing the amplifying gains, and hence one can focus, without loss of optimality, only on the receive sensors' power gains $\pmb{p}=[p_1 \cdots p_N]^T$, with $p_n=|\alpha_n|^2$, for $n \in \mathcal{N}$.   

As in the CF backhaul transmission in Section \ref{sec:CFopt}, we can write the hypotheses (\ref{eq:multi_FCNW}) in a standard form by whitening the signal received at the fusion center, and consequently the detection problem can be expressed as (\ref{eq:ortho_FCW}), where we have redefined $\pmb{y}=\pmb{D}\tilde{\pmb{r}}$; $\pmb{D}=(\sum_{n=1}^{N}(|f_n|^2p_n\sigma_{c,n}^2\pmb{x}\pmb{x}^H+|f_n|^2p_n\pmb{\Omega}_{w,n})+\pmb{\Omega}_z)^{-1/2}$ is the whitening filter with respect to the overall additive noise $\sum_{n=1}^{N}f_n\alpha_n(\pmb{c}_n+\pmb{w}_n)+\pmb{z}$; and $\pmb{S}=\sum_{n=1}^{N}|f_n|^2p_n\sigma_{t,n}^2\pmb{x}\pmb{x}^H$ is the correlation matrix of the desired signal part. Accordingly, the detection problem has the standard estimator-correlator solution given by the test in (\ref{eq:ortho_test}). In the rest of this section, we seek to optimize the detection performance with respect to the waveform $\pmb{x}$ and the power gains $\pmb{p}$, under power constraints on the transmit element and receive sensors. As done above, we adopt the Bhattacharyya distance as the performance metric. As per (\ref{eq:Bdist}), the Bhattacharyya distance between the distributions (\ref{eq:multi_FCNW}) of the signals received at the fusion center under the two hypotheses $\mathcal{H}_0$ and $\mathcal{H}_1$ can be calculated as 
\vspace{-0.2cm}
\begin{eqnarray}\label{eq:multi_Bdist}
\mathcal{B}(\pmb{x}, \pmb{p}; \pmb{f})&=&\log\left(\frac{\left|\pmb{I}+0.5\pmb{D}\pmb{S}\pmb{D}\right|}{\sqrt{\left|\pmb{I}+\pmb{D}\pmb{S}\pmb{D}\right|}}\right)\nonumber\\
&=&\log\left(\frac{1+0.5\lambda}{\sqrt{1+\lambda}}\right),
\end{eqnarray}
where $\lambda=\pmb{f}^H\pmb{P}\pmb{\Sigma}_t\pmb{f}\pmb{x}^H(\pmb{f}^H\pmb{P}\pmb{\Sigma}_c\pmb{f}\pmb{x}\pmb{x}^H+(\pmb{f}\otimes\pmb{I}_K)^H$ $(\pmb{P}\otimes\pmb{I}_K)\pmb{\Omega}_w(\pmb{f}\otimes\pmb{I}_K)+\pmb{\Omega}_z)^{-1}\pmb{x}$; $\pmb{\Sigma}_t=\text{diag}\{\sigma_{t,1}^2,\dots$ $,\sigma_{t,N}^2\}$ and $\pmb{\Sigma}_c=\text{diag}\{\sigma_{c,1}^2, \dots, \sigma_{c,N}^2\}$ are the diagonal matrices whose components are the second-order statistics of channel amplitudes of target return and clutter, respectively; $\pmb{\Omega}_w=\text{diag}\{\pmb{\Omega}_{w,1}, \dots, \pmb{\Omega}_{w,N}\}\in R^{NK\times NK}$ is a block diagonal matrix containing all the noise covariance matrices at the receive sensors; and $\pmb{P}=\text{diag}\{\pmb{p}\} \in R^{N \times N}$ is the diagonal matrix that contains the receive sensors' power gains. Note that we have made explicit the dependence of the Bhattacharyya distance $\mathcal{B}(\pmb{x}, \pmb{p}; \pmb{f})$ on the channels $\pmb{f}$ at the fusion center, as well as on the waveform $\pmb{x}$ and the receive sensors' power gains $\pmb{p}$. 
\subsection{Short-Term Adaptive Design}\label{sec:multi_probShort}
We first consider the case in which design of the waveform $\pmb{x}$ and of the receive sensors' gains $\pmb{p}$ depends on the instantaneous gain of the CSI of the receive sensors-to-fusion center channels $\pmb{f}$. Note that this design requires to modify the solution vector $(\pmb{x},\pmb{p})$ at the time scale at which the channel vector $\pmb{f}$ varies, hence entailing a potentially large feedback overhead from the fusion center to the receive sensors and the transmit element. The problem of maximizing the Bhattacharyya distance (\ref{eq:multi_Bdist}) over the waveform $\pmb{x}$ and the power gains $\pmb{p}$ under the power constraints for transmit element and receive sensors, is stated as
\vspace{-0.2cm}
\begin{subequations}\label{eq:multi_optShort}
\begin{eqnarray}
&&\hspace{-1.5cm}{\mathop {\text{minimize} }\limits_{\pmb{x},\pmb{p}}} \hspace{0.6cm}{\bar{\mathcal{B}}(\pmb{x}, \pmb{p}; \pmb{f})}\label{eq:multi_optShortObj}\\
&& \hspace{-1.1cm}{\rm{s.t.}}\hspace{0.9cm} {\pmb{x}^H\pmb{x} \le P_T,}\label{eq:multi_optShortConst1}\\
&& \hspace{0.25cm} {\pmb{1}^T\pmb{p} \le P_R,}\\
&& \hspace{0.25cm} {p_n \ge 0, \hspace{0.2cm} n \in \mathcal{N},}\label{eq:multi_optShortConst3}
\end{eqnarray}
\end{subequations}
where we have defined $\bar{\mathcal{B}}(\pmb{x}, \pmb{p};\pmb{f})=-\mathcal{B}(\pmb{x}, \pmb{p}; \pmb{f})$ to formulate the problem as the minimization of the negative Bhattacharyya distance $\bar{\mathcal{B}}(\pmb{x}, \pmb{p}; \pmb{f})$. We observe that the problem (\ref{eq:multi_optShort}) may be easily modified to include individual power constraints at the receive sensors, but this is not further explored here. Moreover, the problem (\ref{eq:multi_optShort}) is not a convex program, since the objective function (\ref{eq:multi_optShortObj}) is not convex. 

We propose an algorithm to solve the optimization problem (\ref{eq:multi_optShort}). As in Section \ref{sec:ortho_algo}, due to the difficulty of obtaining a global optimal solution, we develop a descent algorithm, and adopt the BCD method coupled with MM. The proposed algorithm is summarized in Table Algorithm \ref{al2} and further detailed in Appendix B. The complexity of the Algorithm \ref{al2} by using standard convex optimization tool is polynomial in $K$ and $N$ since, at each outer iteration, MM requires to solve the problems (\ref{eq:multi_optShort_code}) and (\ref{eq:multi_optShort_power}), whose sizes of the optimization domains are $K$ and $N$, and numbers of constraints are $1$ and $N+1$, respectively \cite{Luo10SPM, Boyd04Book}.
\renewcommand{\baselinestretch}{0.8}
\begin{algorithm} [h!]
\begin{algorithmic}
\caption{Short-term adaptive design of waveform and amplifier gain (\ref{eq:multi_optShort})} \label{al2}
\State {\textbf{Initialization (outer loop)}}: Initialize $\pmb{x}^{(0)} \in C^{K \times 1}$, $\pmb{p}^{(0)} \succeq 0$ and set $i=0$.
\State {\textbf{Repeat (BCD method)}}
\State \indent $i \gets i+1$
\State \indent {\textbf{Initialization (inner loop)}}: Initialize $\pmb{x}^{(i,0)}=$ 
\State \indent $\pmb{x}^{(i-1)}$ and set $j=0$.
\State \indent {\textbf{Repeat (MM method for $\pmb{x}^{(i)}$)}}
\State \indent \indent $j \gets j+1$
\State \indent \indent Find $\pmb{x}^{(i,j)}$ by solving the problem (\ref{eq:multi_optShort_code}) with 
\State \indent \indent $\pmb{p}=\pmb{p}^{(i-1)}$.
\State \indent {\textbf{Until}} a convergence criterion is satisfied.
\State \indent {\textbf{Update}} $\pmb{x}^{(i)} \gets \pmb{x}^{(i,j)}$
\State \indent {\textbf{Initialization (inner loop)}}: Initialize $\pmb{p}^{(i,0)}=$ 
\State \indent $\pmb{p}^{(i-1)}$ and set $j=0$. 
\State \indent {\textbf{Repeat (MM method for $\pmb{p}^{(i)}$)}}
\State \indent \indent $j \gets j+1$
\State \indent \indent Find $\pmb{p}^{(i,j)}$ by solving the problem (\ref{eq:multi_optShort_power}) with
\State \indent \indent $\pmb{x}=\pmb{x}^{(i)}$.
\State \indent {\textbf{Until}} a convergence criterion is satisfied.
\State \indent {\textbf{Update}} $\pmb{p}^{(i)} \gets \pmb{p}^{(i,j)}$
\State {\textbf{Until}} a convergence criterion is satisfied. 
\State {\textbf{Solution}}: $\pmb{x} \gets \pmb{x}^{(i)}$ and  $\pmb{p} \gets \pmb{p}^{(i)}$
\end{algorithmic}
\end{algorithm}
\renewcommand{\baselinestretch}{1}
\subsection{Long-Term Adaptive Design}\label{sec:multi_probLong}
Here, in order to avoid the possibly excessive feedback overhead between fusion center and the transmit element and receive sensors of the short-term adaptive solution, we adopt the average Bhattacharyya distance, as the performance criterion, where the average is taken with respect to the distribution of the receive sensors-to-fusion center channels $\pmb{f}$. In this way, the waveform $\pmb{x}$ and receive sensors' gains $\pmb{p}$ have to be updated only at the time scale at which the statistics of channels and noise terms vary. Then, the problem for the long-term adaptive design is formulated from problem (\ref{eq:multi_optShort}) by substituting the objective function $\bar{\mathcal{B}}(\pmb{x}, \pmb{p}; \pmb{f})$ with $E_{\pmb{f}}[\bar{\mathcal{B}}(\pmb{x}, \pmb{p}; \pmb{f})]$, yielding
\vspace{-0.2cm}
\begin{subequations}\label{eq:multi_optLong}
\begin{eqnarray}
&&\hspace{-1.5cm}{\mathop {\text{minimize} }\limits_{\pmb{x},\pmb{p}}} \hspace{0.6cm}{E_{\pmb{f}}\left[\bar{\mathcal{B}}(\pmb{x}, \pmb{p}; \pmb{f})\right]}\label{eq:multi_optLongObj}\\
&& \hspace{-1.1cm}{\rm{s.t.}} \hspace{1.1cm} {\text{(\ref{eq:multi_optShortConst1})}-\text{(\ref{eq:multi_optShortConst3})}.}
\end{eqnarray}
\end{subequations}
Note that the problem (\ref{eq:multi_optLong}) is a stochastic program with a non-convex objective function (\ref{eq:multi_optLongObj}).

Since the stochastic program (\ref{eq:multi_optLong}) has a non-convex objective function, we apply the stochastic successive upper-bound minimization method (SSUM) \cite{Luo13Arxiv}, which minimizes at each step an approximate ensemble average of a locally tight upper bound of the cost function. Specifically, we develop a BCD scheme similar to the one detailed in Table Algorithm \ref{al2} that uses SSUM in lieu of the MM scheme. Details are provided in Appendix C. The final algorithm for long-term adaptive design can be summarized as in Table Algorithm $\ref{al2}$ by substituting (\ref{eq:multi_optShort_code}) and (\ref{eq:multi_optShort_power}) with (\ref{eq:multi_optLong_code}) and (\ref{eq:multi_optLong_power}), respectively. Convergence of the SSUM algorithm is proved in \cite{Luo13Arxiv} and the algorithm guarantees feasible iterates. The complexity of the proposed algorithm by using standard convex optimization tool is polynomial in $K$ and $N$ since, at each outer iteration, SSUM requires to solve the problems (\ref{eq:multi_optLong_code}) and (\ref{eq:multi_optLong_power}), whose sizes of the optimization domains are $K$ and $N$, and numbers of constraints are $1$ and $N+1$, respectively \cite{Luo10SPM, Boyd04Book}.
\section{Numerical Results}\label{sec:num}
In the following, the performance of the proposed algorithms that perform joint optimization of the waveform $\pmb{x}$ and of the quantization noise covariance matrices $\pmb{\Omega}_q$ for the CF, and of the waveform $\pmb{x}$ and of the power gains $\pmb{p}$ for AF, are investigated via numerical results in Section \ref{sec:num_ortho} and in Section \ref{sec:num_multi}, respectively. Throughout, we set the length of the waveform to $K=13$ and the variances of the target amplitudes as $\sigma_{t,n}^2=1$ for $n \in \mathcal{N}$. For reference, we consider a baseline waveform with Barker code of length $13$, i.e., $\pmb{b}_{13}=[1\,\,\,\, 1 \,\,\,\, 1 \,\,\,\,1 \,\,\,\,1\,\,-1\,\,$ $-1\,\,\,\, 1\,\,\,\, 1\,\,$ $-1\,\,\,\, 1 \,\,-1 \,\,\,\,1]^T$. Moreover, unless stated otherwise, we model the noise with covariance matrices $[\pmb{\Omega}_{w,n}]_{i,j}=(1-0.12n)^{|i-j|}$ and $[\pmb{\Omega}_z]_{i,j}=(1-0.6)^{|i-j|}$ as in \cite{Stoica13TSP}, hence accounting for temporally correlated interference. The channel coefficients $f_{n}$ have unit variance, i.e., $\sigma_{f_n}^2=1$.
\subsection{CF Backhaul Transmission}\label{sec:num_ortho}
\begin{figure}[t]
\begin{center}
\includegraphics[width=7.5cm, height=6cm]{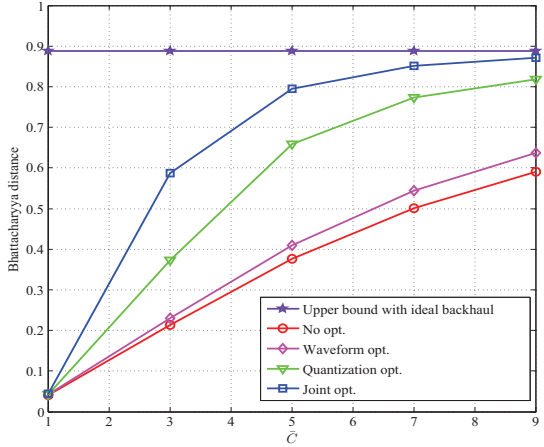}
\caption{Bhattacharyya distance versus the backhaul capacity $\bar{C}_n=\bar{C}$, $n \in \mathcal{N}$ for CF backhaul transmission, with $P_T=10$ dB, $K=13$, $N=3$, $\sigma_{t,n}^2=1$, $\sigma_{c,1}^2=0.125$, $\sigma_{c,2}^2=0.25$, $\sigma_{c,3}^2=0.5$ and $[\pmb{\Omega}_{w,n}]_{i,j}=(1-0.12n)^{|i-j|}$ for $n \in \mathcal{N}$.} \label{fig:BvsC}
\end{center}
\vspace{-0.8cm}
\end{figure}

\begin{figure}[t]
\begin{center}
\includegraphics[width=7.5cm, height=6cm]{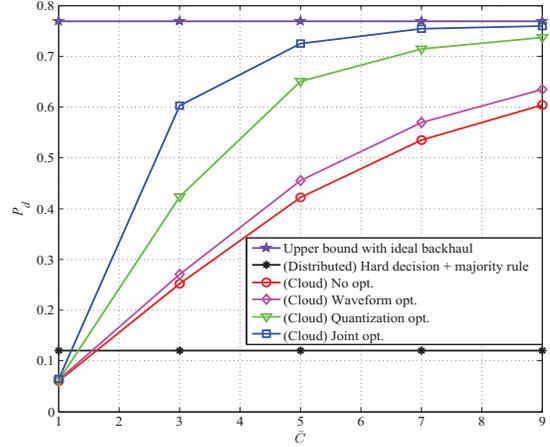}
\caption{Probability of detection $P_d$ versus the backhaul capacity $\bar{C}_n=\bar{C}$ for CF backhaul transmission, $n \in \mathcal{N}$, with $P_T=10$ dB, $K=13$, $N=3$, $\sigma_{t,n}^2=1$, $\sigma_{c,1}^2=0.125$, $\sigma_{c,2}^2=0.25$, $\sigma_{c,3}^2=0.5$, $[\pmb{\Omega}_{w,n}]_{i,j}=(1-0.12n)^{|i-j|}$ for $n \in \mathcal{N}$ and $P_{fa}=0.01$.} \label{fig:PdvsC}
\end{center}
\vspace{-0.8cm}
\end{figure} 

\begin{figure}[t]
\begin{center}
\includegraphics[width=7.5cm, height=6cm]{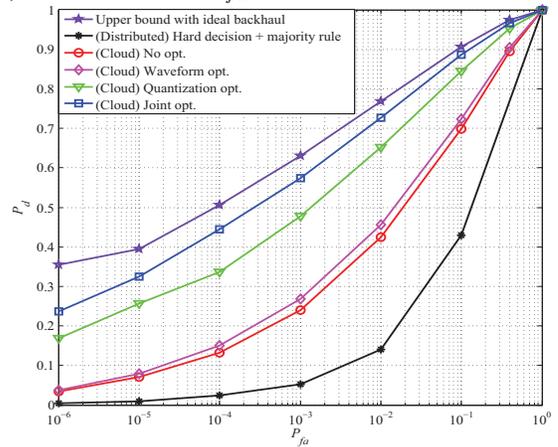}
\caption{ROC curves for CF backhaul transmission with $P_T=10$ dB, $K=13$, $N=3$, $\sigma_{t,n}^2=1$, $\sigma_{c,1}^2=0.125$, $\sigma_{c,2}^2=0.25$, $\sigma_{c,3}^2=0.5$, $[\pmb{\Omega}_{w,n}]_{i,j}=(1-0.12n)^{|i-j|}$ and $\bar{C}_n=\bar{C}=5$ for $n \in \mathcal{N}$.} \label{fig:ROC_ortho}
\end{center}
\vspace{-1cm}
\end{figure}

In this section, the performance of the proposed joint optimization of the waveform $\pmb{x}$ and of the quantization noise covariance matrices $\pmb{\Omega}_q$ in Section \ref{sec:CFopt} is verified via numerical results. Note that some limited results for a sum-backhaul constraint were presented in \cite{Osvaldo14TSPLett}. For reference, we consider the performance of the upper bound obtained with infinite capacity backhaul links, distributed detection using the Barker waveform (see, Section \ref{sec:dist}), and the following strategies: (\textit{i}) \textit{No optimization  (No opt.)}: Set $\pmb{x}=\sqrt{P_T/K}\pmb{b}_{13}$ and $\pmb{\Omega}_{q,n}=\epsilon\pmb{I}$, for $n \in \mathcal{N}$, where $\epsilon$ is a constant that is found by satisfying the constraint (\ref{eq:ortho_optRate}) with equality; (\textit{ii}) \textit{Waveform optimization (Waveform opt.)} : Optimize the waveform $\pmb{x}$ by using the algorithm in \cite{Stoica13TSP}, which is given in Algorithm \ref{al1} by setting $\pmb{\Omega}_{q,n}=0$ for $n \in \mathcal{N}$, and set $\pmb{\Omega}_{q,n}=\epsilon\pmb{I}$, for $n \in \mathcal{N}$, as explained above; (\textit{iii}) \textit{Quantization noise optimization (Quantization opt.)}: Optimize the covariance matrices $\pmb{\Omega}_q$ as per Algorithm \ref{al1} with $\pmb{x}=\sqrt{P_T/K}\pmb{b}_{13}$. In the following, we set the number of receive sensors, the transmit power and the variance of the clutter amplitudes as $N=3$, $P_T=10$ dB, $\sigma_{c,1}^2=0.125$, $\sigma_{c,2}^2=0.25$ and $\sigma_{c,3}^2=0.5$, respectively. Also, the backhaul rate constraints $\bar{C}_n$ are assumed to be equal, i.e., $\bar{C}_n=\bar{C}$ for all $n \in \mathcal{N}$. 

\begin{figure*}[t]
\centering
\subfigure[Low-frequency interference]{
    \includegraphics[width=7.5cm, height=6cm]{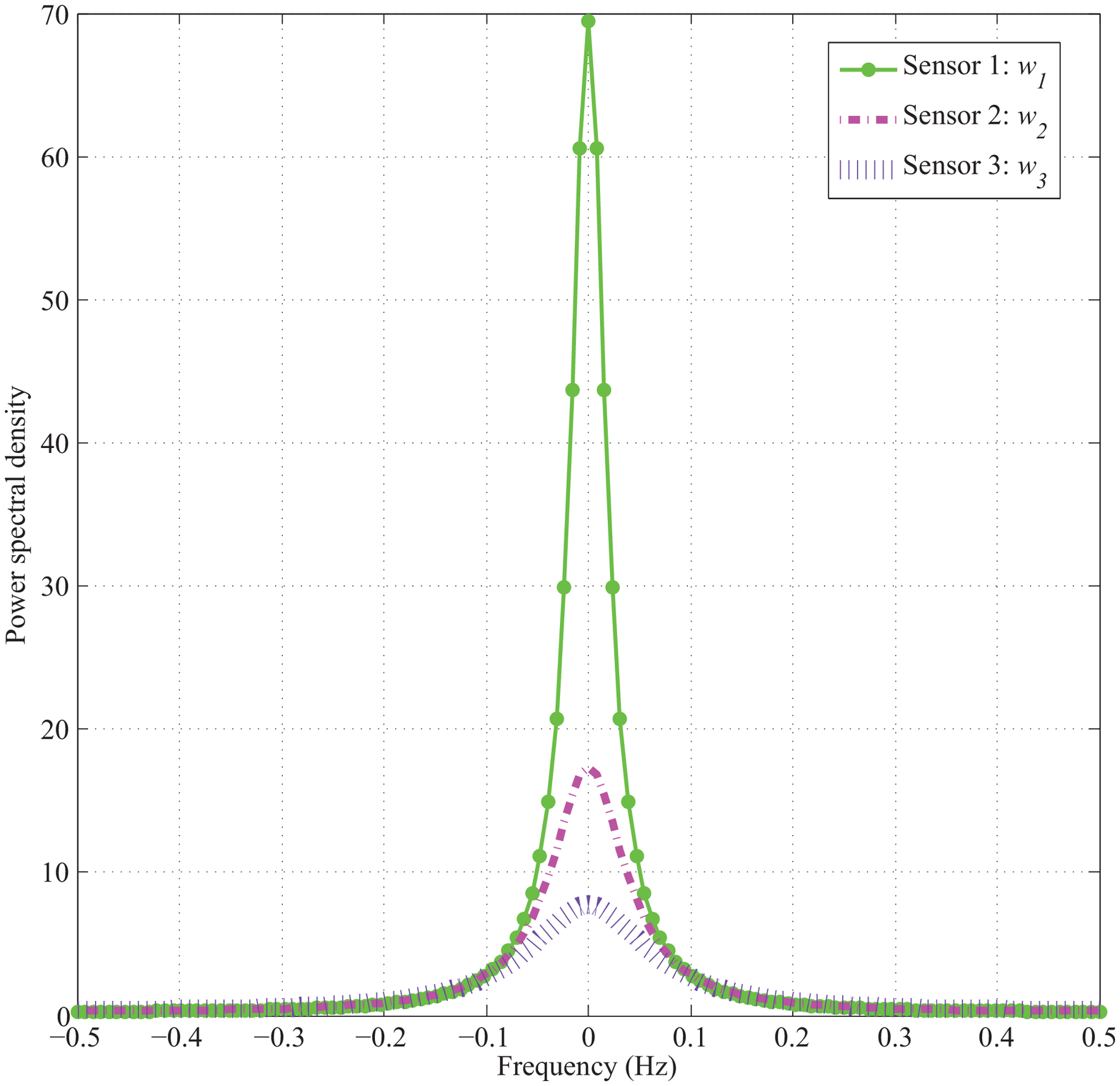}
    \label{fig:subfig1}
}
\subfigure[High-frequency interference]{
   \includegraphics[width=7.5cm, height=6cm]{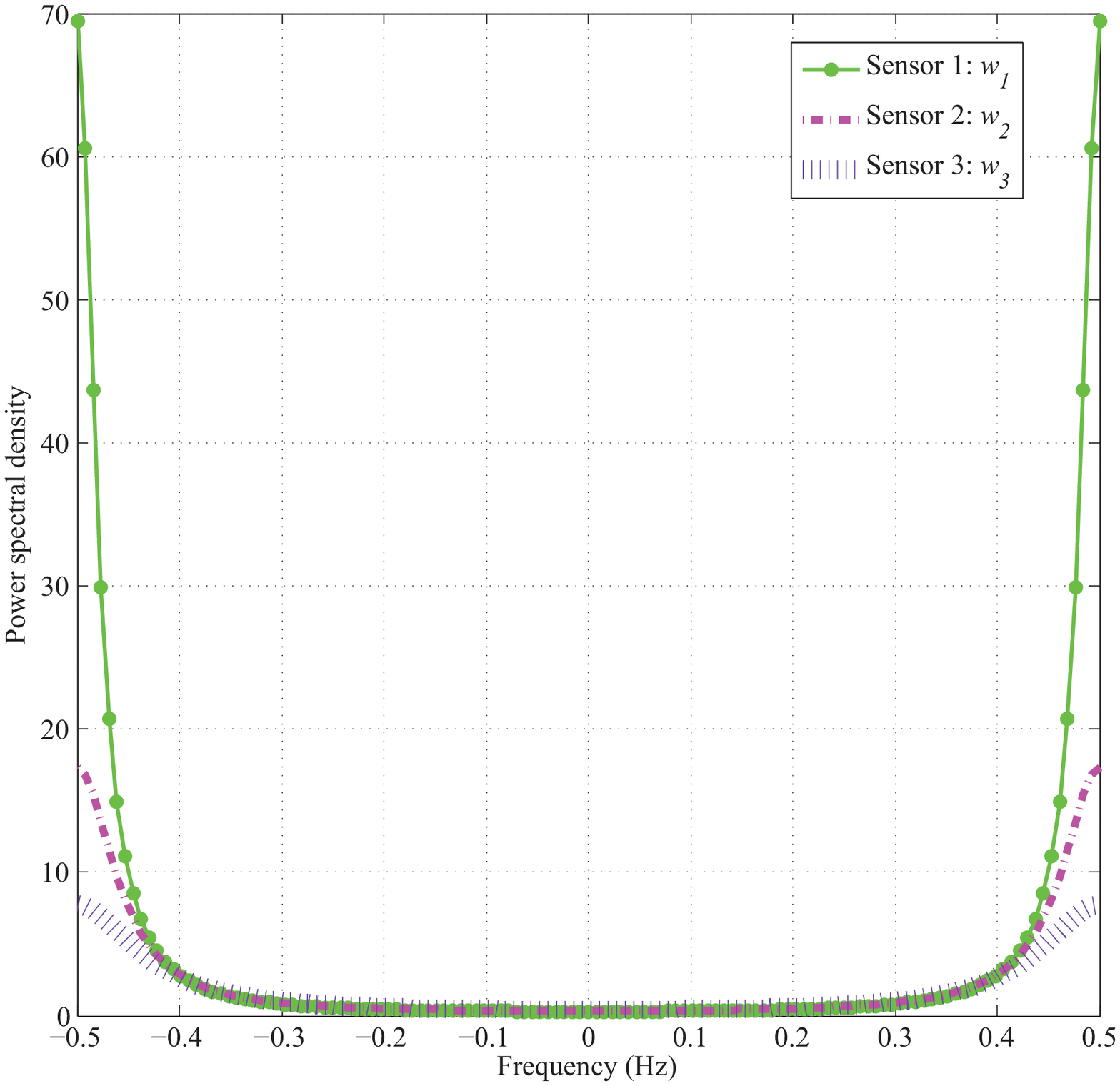}
    \label{fig:subfig2} 
}
\subfigure[Optimal waveform and quantization noise with low-frequency interference]{
    \includegraphics[width=7.5cm, height=6cm]{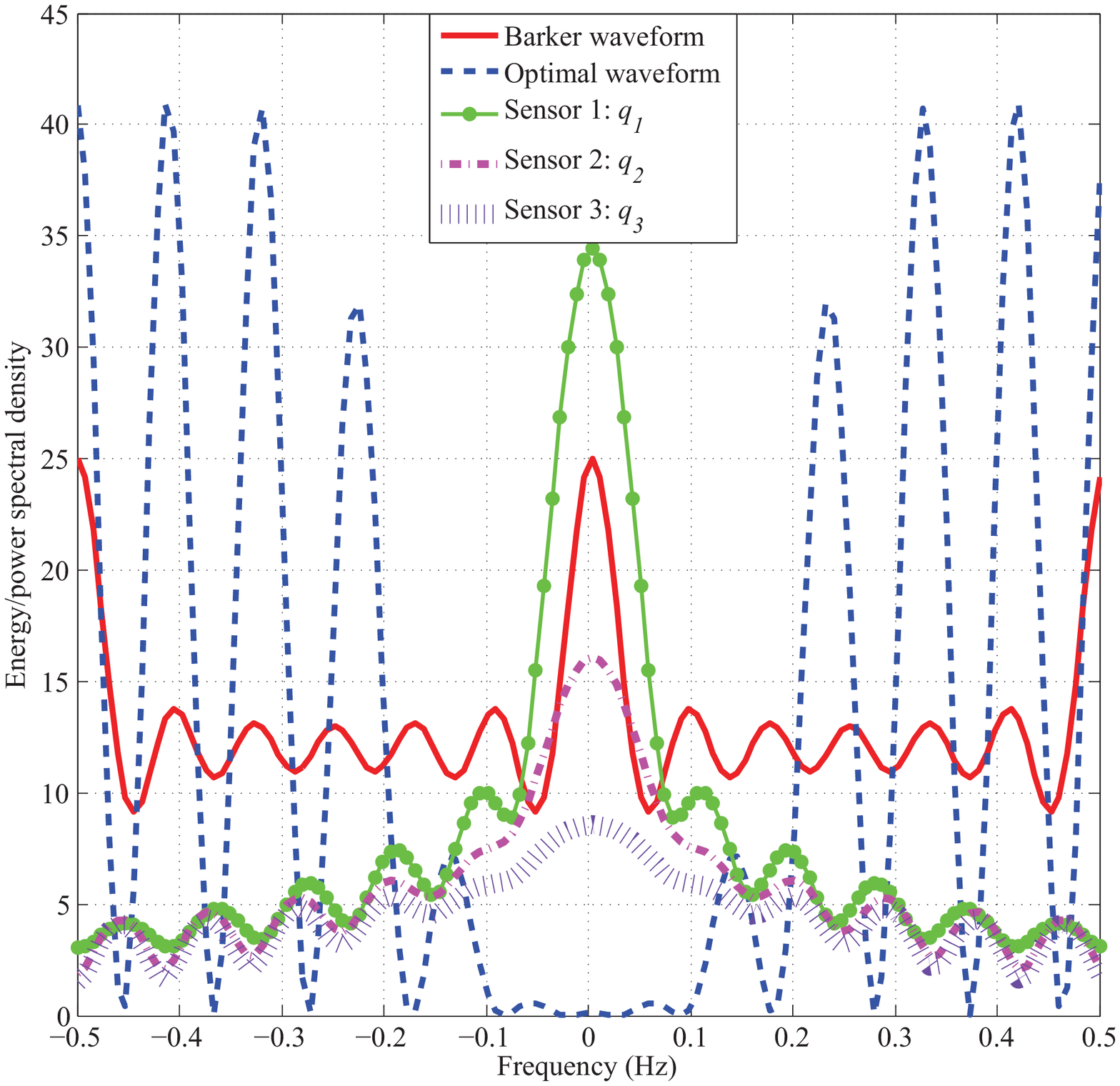}
    \label{fig:subfig3}

}
\subfigure[Optimal waveform and quantization noise with high-frequency interference]{
    \includegraphics[width=7.5cm, height=6cm]{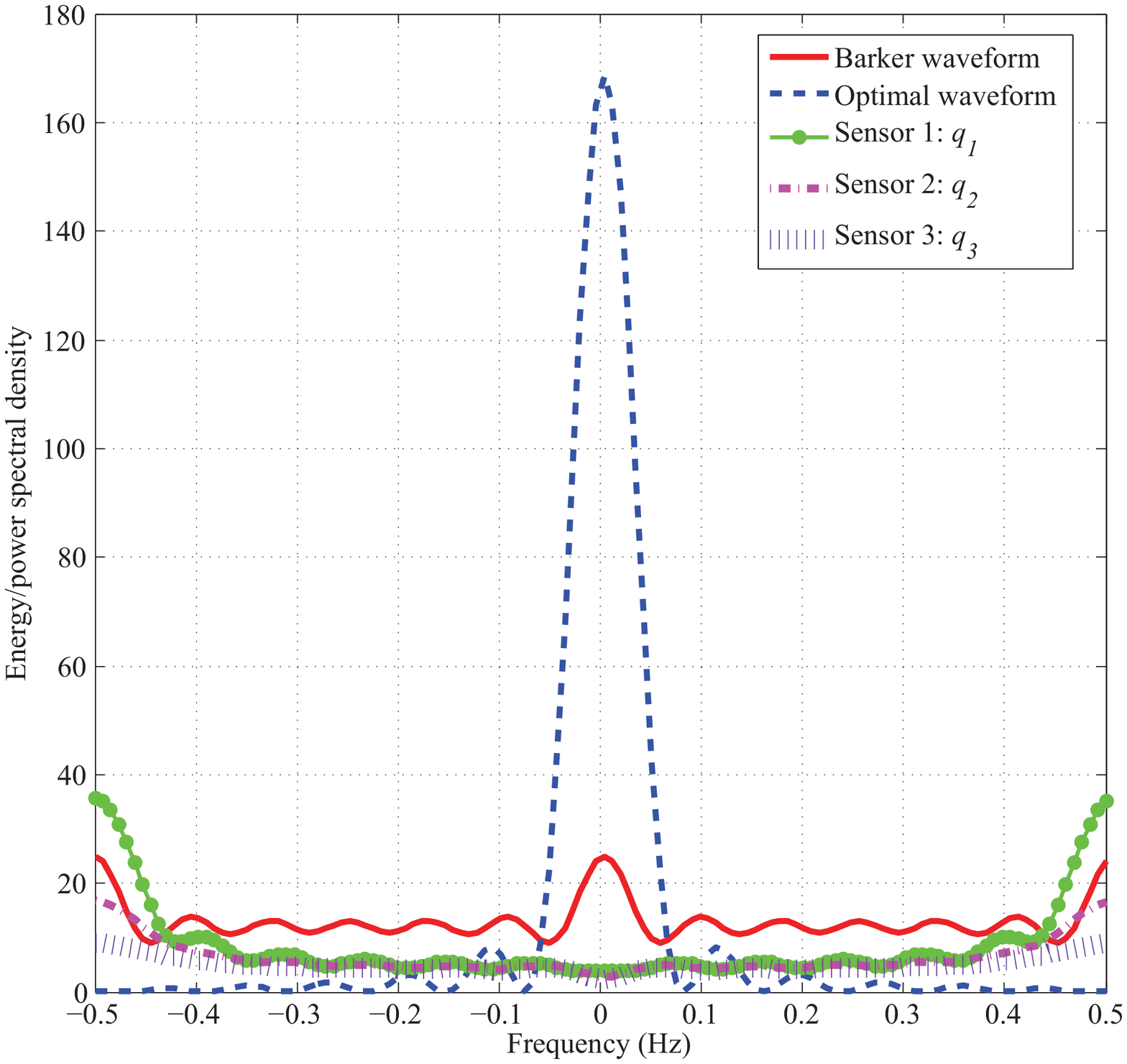}
    \label{fig:subfig4}
}
\caption{Comparison of the energy/power spectral densities of the waveforms obtained with a Barker code (Barker waveform) and with an optimal code $\pmb{x}$ (Optimal waveform), and of optimal quantization noise $\{\pmb{q}_n\}_{n=1}^N$ obtained by Algorithm \ref{al1} when $P_T=10$ dB, $K=13$, $N=3$, $\sigma_{t,n}^2=1$, $\sigma_{c,1}^2=0.125$, $\sigma_{c,2}^2=0.25$, $\sigma_{c,3}^2=0.5$, and $\bar{C}_n=\bar{C}=5$ for $n \in \mathcal{N}$: (a) and (c) consider receive sensors with low-frequency interference having temporal correlation $[\pmb{\Omega}_{w,n}]_{i,j}=(1-0.12n)^{|i-j|}$, (b) and (d) consider receive sensors with high-frequency interference having temporal correlation $[\pmb{\Omega}_{w,n}]_{i,j}=(-1+0.12n)^{|i-j|}$.}
\label{fig:orthoPSD}
\vspace{-0.5cm}
\end{figure*}

In Fig. \ref{fig:BvsC} the Bhattacharyya distance is plotted versus the available backhaul capacity $\bar{C}$. For intermediate and large values of $\bar{C}$, the proposed joint optimization of waveform and quantization noise is seen to be significantly beneficial over all separate optimization strategies. In order to study the actual detection performance and validate the results in Fig. \ref{fig:BvsC}, Fig. \ref{fig:PdvsC} shows the detection probability $P_d$ as a function of the available backhaul capacity $\bar{C}$ when the false alarm probability is $P_{fa}=0.01$. The curve was evaluated via Monte Carlo simulations by implementing the optimum test detector (\ref{eq:ortho_test}). We also implemented the distributed detection scheme described in Section \ref{sec:dist} by setting the threshold $\gamma_n$ to be equal for $n \in \mathcal{N}$ for simplicity. It can be noted that the relative gains predicted by the Bhattacharyya distance criterion in Fig. \ref{fig:BvsC} are consistent with the performance shown in Fig. \ref{fig:PdvsC}. Moreover, for small values of $\bar{C}$, distributed detection outperforms cloud detection due to the performance degradation caused by the large quantization noise on the cloud-based schemes. However, as the available backhaul capacity $\bar{C}$ increases, the cloud detection approach considerably outperforms distributed detection.  

Fig. \ref{fig:ROC_ortho} plots the Receiving Operating Characteristic (ROC), i.e., the detection probability $P_d$ versus false alarm probability $P_{fa}$, for $\bar{C}=5$. It is confirmed that the proposed joint optimization method provides remarkable gains over all separate optimization schemes as well as over the distributed detection approach. For instance, for $P_{fa}=0.01$, joint optimization yields $P_d=0.7251$, while waveform optimization only yields $P_d=0.4556$.

Fig. \ref{fig:orthoPSD} shows the energy/power spectral density functions of the waveform with Barker code \textit{(Barker waveform)} and with optimal code $\pmb{x}$ \textit{(Optimal waveform)}, and of optimal quantization noise $\{\pmb{q}_n\}_{n=1}^N$ obtained by Algorithm \ref{al1} when a square root Nyquist chip waveform $\phi(t)$ with duration $T_c$ is adopted, and $\bar{C}_n=\bar{C}=5$ for $n \in \mathcal{N}$. We consider two types of interference at the receive sensors, namely (a) low-frequency interference with temporal correlation  $[\pmb{\Omega}_{w,n}]_{i,j}=(1-0.12n)^{|i-j|}$ which has a single spectral peak at zero frequency; and (b) high-frequency interference with  temporal correlation $[\pmb{\Omega}_{w,n}]_{i,j}=(-1+0.12n)^{|i-j|}$, having a minimum at zero frequency. It is observed in Fig. \ref{fig:subfig3} and Fig. \ref{fig:subfig4} that the spectrum of the optimal waveform concentrates the transmitted energy at frequencies for which the interference power is less pronounced, while the spectrum of the quantization noise concentrates at frequencies and sensors for which the interference power is more pronounced.  
\subsection{AF Backhaul Transmission}\label{sec:num_multi}
In this section, we evaluate the performance of the proposed algorithms that perform the joint optimization of the waveform $\pmb{x}$ and of the amplifying power gains $\pmb{p}$ for the short-term (Section \ref{sec:multi_probShort}) and long-term (Section \ref{sec:multi_probLong}) adaptive designs. For reference, we consider the following schemes; \textit{(i) No opt.}: Set $\pmb{x}=\sqrt{P_T/K}\pmb{b}_{13}$ and $\pmb{p}=P_R/N\pmb{1}_N$; \textit{(ii) Waveform opt.}: Optimize the waveform $\pmb{x}$ as per Algorithm $\ref{al2}$ (with (\ref{eq:multi_optLong_code}) in lieu of (\ref{eq:multi_optShort_code}) for the long-term adaptive design) with $\pmb{p}=P_R/N\pmb{1}_N$; and \textit{(iii) Gain optimization (Gain opt.)}: Optimize the gains $\pmb{p}$ as per Algorithm $\ref{al2}$ (with (\ref{eq:multi_optLong_power}) in lieu of (\ref{eq:multi_optShort_power}) for the long-term adaptive design) with $\pmb{x}=\sqrt{P_T/K}\pmb{b}_{13}$. We set the total receive sensors' power as $P_R=10$ dB. Note that the upper bound with ideal backhaul is far from the performance achieved with AF over a non-orthogonal backhaul even for large sensors' power $P_R$, and it is hence not shown here. The gap between the AF performance and the upper bound is due to the fact that, in order to obtain an ideal backhaul, one needs to code across long block lengths whereas AF operates on block length of size equal to the waveform $K$ (here $K=13$).

\begin{figure}[t]
\begin{center}
\includegraphics[width=7.5cm, height=6cm]{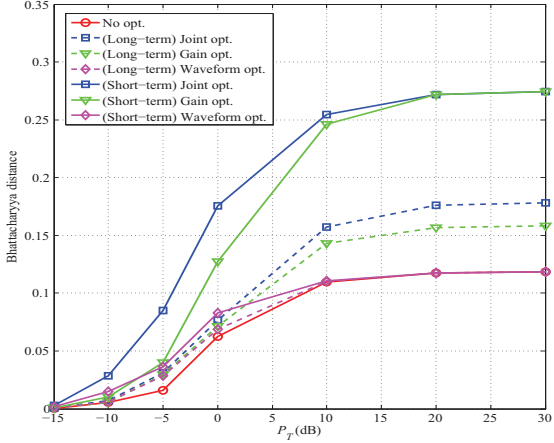}
\caption{Bhattacharyya distance versus the transmit element's power $P_T$ for AF backhaul transmission with $P_R=10$ dB, $K=13$, $N=3$, $\sigma_{f_n}^2=1$, $\sigma_{t,n}^2=1$, $\sigma_{c,1}^2=0.25$, $\sigma_{c,2}^2=0.5$, $\sigma_{c,3}^2=1$, $[\pmb{\Omega}_{w,n}]_{i,j}=(1-0.12n)^{|i-j|}$ and $[\pmb{\Omega}_z]_{i,j}=(1-0.6)^{|i-j|}$ for $n \in \mathcal{N}$.} \label{fig:PT}
\end{center}
\vspace{-0.5cm}
\end{figure}
 
\begin{figure}[t]
\begin{center}
\includegraphics[width=7.5cm, height=6cm]{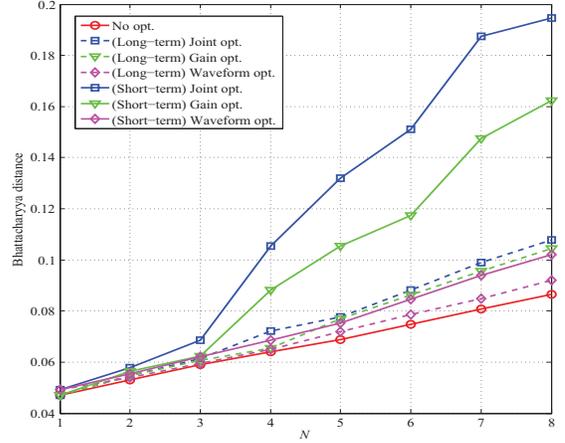}
\caption{Bhattacharyya distance versus the number receive sensors $N$ for AF backhaul transmission with $P_T=5$ dB, $P_R=10$ dB, $K=13$, $\sigma_{f_n}^2=1$, $\sigma_{t,n}^2=1$, $\sigma_{c,1}^2=1$, $\sigma_{c,2}^2=0.9$, $\sigma_{c,3}^2=0.75$, $\sigma_{c,4}^2=0.5$, $\sigma_{c,5}^2=0.35$, $\sigma_{c,6}^2=0.25$ and $\sigma_{c,7}^2=0.125$, $\sigma_{c,8}^2=0.05$, $[\pmb{\Omega}_{w,n}]_{i,j}=(1-0.12n)^{|i-j|}$ and $[\pmb{\Omega}_z]_{i,j}=(1-0.6)^{|i-j|}$ for $n \in \mathcal{N}$.}
\label{fig:N}
\end{center}
\vspace{-0.5cm}
\end{figure} 

\begin{figure}[t]
\begin{center}
\includegraphics[width=7.5cm, height=6cm]{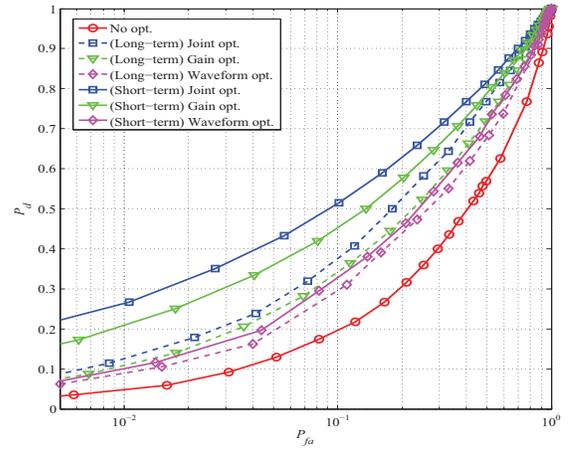}
\caption{ROC curves for AF backhaul transmission with $P_T=5$ dB, $P_R=10$ dB, $K=13$, $N=3$, $\sigma_{f_n}^2=1$, $\sigma_{t,n}^2=1$, $\sigma_{c,1}^2=0.25$, $\sigma_{c,2}^2=0.5$, $\sigma_{c,3}^2=1$, $[\pmb{\Omega}_{w,n}]_{i,j}=(1-0.12n)^{|i-j|}$ and $[\pmb{\Omega}_z]_{i,j}=(1-0.6)^{|i-j|}$ for $n \in \mathcal{N}$.} \label{fig:ROC_multi}
\end{center}
\vspace{-0.5cm}
\end{figure}
Fig. \ref{fig:PT} shows the Bhattacharyya distance as a function of the transmit element's power $P_T$, with $N=3$, $\sigma_{c,1}^2=0.25$, $\sigma_{c,2}^2=0.5$ and $\sigma_{c,3}^2=1$. For small values of $P_T$, optimizing the waveform is more advantageous than optimizing the amplifying gains, due to the fact that performance is limited by the transmit element-to-receive sensors connection. In contrast, for intermediate and large values of $P_T$, the optimization of the receive sensors' gains is to be preferred, since the performance becomes limited by the channels between the receive sensors and the fusion center. Joint optimization significantly outperforms all other schemes, except in the very low- and large-power regimes, in which, as discussed, the performance is limited by either the transmit element-to-receive sensors or the receive sensors-to-fusion center channels. In addition, we observe that the long-term adaptive scheme loses about $30\%$ in terms of the Bhattacharyya distance with respect to the short-term adaptive design in the high SNR regime. The results in Fig. \ref{fig:PT} can be interpreted by noting that the joint optimization seeks to design the transmitted signal $\pmb{x}$ such that it reduces the power transmitted at the frequencies in which the receive sensors observe the largest interference, while, at the same time, allocating more power to receive sensors suffering from less interference and, with the short-term adaptive design, having better channels to the fusion center. 

In Fig. \ref{fig:N}, the Bhattacharyya distance is plotted versus the number receive sensors $N$ with $P_T=5$ dB, $\sigma_{c,1}^2=1$, $\sigma_{c,2}^2=0.9$, $\sigma_{c,3}^2=0.75$, $\sigma_{c,4}^2=0.5$, $\sigma_{c,5}^2=0.35$, $\sigma_{c,6}^2=0.25$, $\sigma_{c,7}^2=0.125$ and $\sigma_{c,8}^2=0.05$. Optimizing the receive sensors' power gains is seen to be especially beneficial at large $N$, due to the ability to allocate more power to the receive sensors in better condition in terms of interference and channels to the fusion center. For instance, even with the long-term adaptive design, optimizing the receive sensors' power gains outperforms waveform optimization with short-term adaptive design for sufficiently large $N$.

Fig. \ref{fig:ROC_multi} plots the ROC curves with $P_T=5$ dB, $N=3$, $\sigma_{c,1}^2=0.25$, $\sigma_{c,2}^2=0.5$ and $\sigma_{c,3}^2=1$. The curve was evaluated via Monte Carlo simulations by implementing the optimum test detector (\ref{eq:ortho_test}) as discussed in Section \ref{sec:AFopt}. It can be observed that the gains observed in the previous figures directly translate into a better ROC performance of joint optimization. Note also that power gain optimization is seen to be advantageous due to sufficient value of $P_T$ as predicted based on Fig. \ref{fig:PT}.   
\section{Concluding Remarks}\label{sec:con}
We have studied a multistatic cloud radar system, where the receive sensors and fusion center are connected via an orthogonal-access backhaul or a non-orthogonal multiple-access backhaul channel. In the former case, each receive sensor quantizes and forwards the signal sent by transmit element to a fusion center following a compress-and-forward protocol, while amplify-and-forward of the received signal is carried out over the multiple-access backhaul. The fusion center collects the signals from all the receive sensors and determines the target's presence or absence. We have investigated the joint optimization of waveform and backhaul transmission so as to maximize the detection performance. As the performance metric, we adopted the Bhattacharyya distance and the proposed algorithmic solutions were based on successive convex approximations. Overall, joint optimization was seen to have remarkable gains over the standard separate optimization of waveform and backhaul transmission. Moreover, cloud processing is found to outperform the standard distributed detection approach as long as the backhaul capacity is large enough.
\begin{appendices}
\section*{Appendix A: Details of CF Optimization}  
\renewcommand{\theequation}{A.\arabic{equation}}
\setcounter{equation}{0}  
\renewcommand{\thesection}{A}
\subsection{Review of MM Method}\label{sec:MM}
We start by reviewing the MM method. For a non-convex function $f(\pmb{t})$ of a generic variable $\pmb{t}$, which may appear either in the cost function or among the constraints, the MM method substitutes at the $l$th iteration, a convex approximation $f(\pmb{t}|\pmb{t}^{(l-1)})$ of $f(\pmb{t})$, such that the global upper bound property $f(\pmb{t}|\pmb{t}^{(l-1)})\ge f(\pmb{t})$ is satisfied for all $\pmb{t}$ in the domain, along with the local tightness condition $f(\pmb{t}^{(l-1)}|\pmb{t}^{(l-1)})=f(\pmb{t}^{(l-1)})$. These properties guarantee the feasibility of all iterates and the descent property that the object function does not increase along the iterations.  
\subsection{Details of the Proposed Algorithm \ref{al1}}\label{sec:ortho_algo_pro}
In the following, we discuss the application of the MM method to perform optimizations over $\pmb{x}$ and $\pmb{\Omega}_q$ in Algorithm \ref{al1}, respectively.

{\bf{Optimization over $\pmb{x}$:}} Here, the goal is to obtain the optimal value of $\pmb{x}^{(i)}$ for problem (\ref{eq:ortho_opt}) given $\pmb{\Omega}_q=\pmb{\Omega}_q^{(i-1)}$. To this end, we apply the MM method. Specifically, at the $j$th iteration of the MM method and the $i$th iteration of the outer loop, the MM method solves a QCQP and obtains a solution $\pmb{x}^{(i,j)}$ by substituting the non-convex objective function $\bar{\mathcal{B}}(\pmb{x}, \pmb{\Omega}_q)$ with a tight upper bound $\mathcal{U}^{\bar{\mathcal{B}}}(\pmb{x},\pmb{\Omega}_q|\pmb{x}^{(i,j-1)})$ around the current iterate $\pmb{x}^{(i,j-1)}$. This bound is obtained by linearizing the difference-of-convex functions in $\bar{\mathcal{B}}(\pmb{x}, \pmb{\Omega}_q)$ via the first-order Taylor approximation \cite{Hunter04Amer}, which follows the same steps as in \cite[eq. (34) and (50) in Section IV]{Stoica13TSP}, and is given by
\begin{eqnarray}\label{eq:ortho_code_Bdist}
&&\hspace{-0.9cm}\mathcal{U}^{\bar{\mathcal{B}}}(\pmb{x},\pmb{\Omega}_{q}|\pmb{x}^{(i,j-1)})=\sum_{n=1}^{N}\mathcal{U}^{\bar{\mathcal{B}}}_n(\pmb{x},\pmb{\Omega}_{q,n}|\pmb{x}^{(i,j-1)})\nonumber\\
&&\hspace{-0.9cm}=\sum_{n=1}^N\phi_{n}^{(i,j-1)}\pmb{x}^{H}\left(\pmb{\Omega}_{w,n}+\pmb{\Omega}_{q,n}\right)^{-1}\pmb{x}\nonumber\\
&&\hspace{1.4cm}-\mathfrak{Re}\left(\left(\pmb{d}_{n}^{(i,j-1)}\right)^{H}\pmb{x}\right),\\
&&\hspace{-1.2cm}\text{where}\nonumber
\end{eqnarray}
\begin{eqnarray}
&&\hspace{-0.9cm}\phi_n^{(i,j-1)}=\frac{\beta_n}{1+\beta_n y_n^{(i,j-1)}}+\beta_n(1+0.5\gamma_n)\nonumber\\
&&\hspace{1.4cm}+\frac{0.5\gamma_n}{1+\lambda_n^{(i,j-1)}}\frac{\beta_n}{\left(1+\beta_n y_n^{(i,j-1)}\right)^2};\nonumber\\
&&\hspace{-0.9cm}\pmb{d}_n^{(i,j-1)}=\left(\frac{2\beta\left(1+0.5\gamma_n\right)}{1+\beta_n y_n^{(i,j-1)}\left(1+0.5\gamma_n\right)}\right.\nonumber\\
&&\hspace{-0.2cm}\left.+2\beta_n\left(1+0.5\gamma_n\right)\right) \left(\pmb{\Omega}_{w,n}+\pmb{\Omega}_{q,n}\right)^{-1}\pmb{x}^{(i,j-1)};\nonumber\\
&&\hspace{-0.9cm}y_n^{(i,j-1)}=\left(\pmb{x}^{(i,j-1)}\right)^H\left(\pmb{\Omega}_{w,n}+\pmb{\Omega}_{q,n}\right)^{-1}\pmb{x}^{(i,j-1)};\nonumber\\
&&\hspace{-0.9cm}\lambda_n^{(i,j-1)}=\gamma_n-\frac{\gamma_n}{1+\beta_n y_n^{(i,j-1)}}.\nonumber
\end{eqnarray}
with $\beta_n=\sigma_{c,n}^2$ and $\gamma_n=\sigma_{t,n}^2/\beta_n$.
A bound with the desired property can also be easily derived for $\mathcal{I}_{n}(\pmb{x},\pmb{\Omega}_{q,n})$ by using the inequality $\log(1+t)\le\log(1+t^{(l)})+1/(1+t^{(l)})(t-t^{(l)})$,
for $t=(\sigma_{t,n}^{2}+\sigma_{c,n}^{2})\pmb{x}^{H}(\pmb{\Omega}_{w,n}+\pmb{\Omega}_{q,n})^{-1}\pmb{x}$,
leading to
\vspace{-0.2cm}
\begin{eqnarray}\label{eq:ortho_code_rate}
&&\hspace{-1cm}\mathcal{U}^{\mathcal{I}}_{n}(\pmb{x},\pmb{\Omega}_{q,n}|\pmb{x}^{(i,j-1)})=\log\left|\pmb{I}+(\pmb{\Omega}_{q,n})^{-1}\pmb{\Omega}_{w,n}\right|\nonumber\\
&&\hspace{-0.7cm}+\log(1+t^{(i,j-1)})+\frac{1}{1+t^{(i,j-1)}}\left((\sigma_{c,n}^{2}+\sigma_{t,n}^{2})\right.\nonumber\\
&&\hspace{0.8cm}\left.\pmb{x}^{H}(\pmb{\Omega}_{w,n}+\pmb{\Omega}_{q,n})^{-1}\pmb{x}-t^{(i,j-1)}\right).
\end{eqnarray}
At the $j$th iteration of the MM method and the $i$th outer loop, we evaluate the new iterate $\pmb{x}^{(i,j)}$ by solving the following QCQP problem 
\vspace{-0.2cm}
\begin{subequations}\label{eq:ortho_opt_code}
\begin{eqnarray}
&&\hspace{-0.9cm}\pmb{x}^{(i,j)} \leftarrow {\mathop {\text{argmin} }\limits_{\pmb{x}}} \hspace{0.2cm}{\mathcal{U}^{\bar{\mathcal{B}}}(\pmb{x},\pmb{\Omega}_q|\pmb{x}^{(i,j-1)})}\\
&& \hspace{-1cm}{\rm{s.t.}}\hspace{0.4cm} {\mathcal{U}^{\mathcal{I}}_{n}(\pmb{x},\pmb{\Omega}_{q,n}|\pmb{x}^{(i,j-1)}) \le \bar{C}_n, \hspace{0.2cm}n \in \mathcal{N},}\\
&& \hspace{-0.2cm} {\pmb{x}^H\pmb{x} \le P_T.}
\end{eqnarray}
\end{subequations}
The MM method obtains the solution $\pmb{x}^{(i)}$ for the $i$th iteration of the outer loop by solving the problem (\ref{eq:ortho_opt_code}) iteratively over $j$ until a convergence criterion is satisfied. 

{\bf{Optimization over $\pmb{\Omega}_q$:}} In this part, we consider the optimization of matrices $\pmb{\Omega}_{q}^{(i)}$ for a given $\pmb{x}=\pmb{x}^{(i)}$. Similar to the optimization over $\pmb{x}^{(i)}$, we use upper bounds of $\bar{\mathcal{B}}(\pmb{x}, \pmb{\Omega}_q)$ and $\mathcal{I}_n(\pmb{x}, \pmb{\Omega}_{q,n})$ for optimization. First, by rewriting $\mathcal{I}_n(\pmb{x}, \pmb{\Omega}_{q,n})$ as $\mathcal{I}_{n}(\pmb{x},\pmb{\Omega}_{q,n})=\log|\pmb{\Omega}_{q,n}+(\sigma_{t,n}^{2}+\sigma_{c,n}^{2})\pmb{x}\pmb{x}^{H}+\pmb{\Omega}_{w,n}|-\log|\pmb{\Omega}_{q,n}|$, we obtain difference-of-convex functions with respect to $\pmb{\Omega}_{q,n}$. Then, by linearizing negative convex component via its first-order Taylor approximation, upper bounds $\mathcal{U}^{\mathcal{I}}_{n}(\pmb{x},\pmb{\Omega}_{q,n}|\pmb{\Omega}_{q,n}^{(i,j-1)})$ and $\mathcal{U}^{\bar{\mathcal{B}}}(\pmb{x},\pmb{\Omega}_q|\pmb{\Omega}_q^{(i,j-1)})$ with the desired properties of MM method are derived for functions $\mathcal{I}_{n}(\pmb{x},\pmb{\Omega}_{q,n})$ and $\bar{\mathcal{B}}(\pmb{x},\pmb{\Omega}_q)$, respectively, as follows:
\vspace{-0.2cm}
\begin{eqnarray}\label{eq:ortho_quant_rate}
&&\hspace{-0.7cm}\mathcal{U}^{\mathcal{I}}_{n}(\pmb{x},\pmb{\Omega}_{q,n}|\pmb{\Omega}_{q,n}^{(i,j-1)})\nonumber\\
&&\hspace{-0.7cm}=\log|\pmb{\Omega}_{q,n}^{(i,j-1)}+(\sigma_{t,n}^{2}+\sigma_{c,n}^{2})\pmb{x}\pmb{x}^{H}+\pmb{\Omega}_{w,n}|\nonumber\\
&&\hspace{-0.2cm}-\log|\pmb{\Omega}_{q,n}|+\text{tr}\left\{\left(\pmb{\Omega}_{q,n}^{(i,j-1)}+(\sigma_{t,n}^{2}+\sigma_{c,n}^{2})\pmb{x}\pmb{x}^{H}\right.\right.\nonumber\\
&&\hspace{1.5cm}\left.\left.+\pmb{\Omega}_{w,n}\right)^{-1}\left(\pmb{\Omega}_{q,n}-\pmb{\Omega}_{q,n}^{(i,j-1)}\right)\right\}
\end{eqnarray}
and
\begin{eqnarray}\label{eq:ortho_quant_Bdist}
&&\hspace{-0.7cm}\mathcal{U}^{\bar{\mathcal{B}}}(\pmb{x},\pmb{\Omega}_q|\pmb{\Omega}_q^{(i,j-1)})=\sum_{n=1}^N\mathcal{U}^{\bar{\mathcal{B}}}_n(\pmb{x},\pmb{\Omega}_{q,n}|\pmb{\Omega}_{q,n}^{(i,j-1)})\nonumber\\
&&\hspace{-0.7cm}=\sum_{n=1}^N-\log|(0.5\sigma_{t,n}^{2}+\sigma_{c,n}^{2})\pmb{x}\pmb{x}^{H}+\pmb{\Omega}_{w,n}+\pmb{\Omega}_{q,n}|\nonumber\\ &&\hspace{-0.7cm}+0.5\text{tr}\left\{\left((\sigma_{t,n}^{2}+\sigma_{c,n}^{2})\pmb{x}\pmb{x}^{H}+\pmb{\Omega}_{w,n}+\pmb{\Omega}_{q,n}^{(i,j-1)}\right)^{-1}\right.\nonumber\\
&&\hspace{-0.7cm}\left.\times\pmb{\Omega}_{q,n}\right\}+0.5\text{tr}\left\{\left(\sigma_{c,n}^{2}\pmb{x}\pmb{x}^{H}+\pmb{\Omega}_{w,n}+\pmb{\Omega}_{q,n}^{(i,j-1)}\right)^{-1}\right.\nonumber\\
&&\hspace{4.7cm}\left.\times\pmb{\Omega}_{q,n}\right\}.
\end{eqnarray}
The $j$th iteration of the MM method then evaluates the matrices $\pmb{\Omega}_q^{(i,j)}=\{\pmb{\Omega}_{q,n}^{(i,j)}\}_{n \in \mathcal{N}}$ by solving the following convex optimization problem
\begin{subequations}\label{eq:ortho_opt_quant}
\begin{eqnarray}
&&\hspace{-0.9cm}\pmb{\Omega}_q^{(i,j)} \leftarrow {\mathop {\text{argmin} }\limits_{\pmb{\Omega}_q}} \hspace{0.1cm}{\mathcal{U}^{\bar{\mathcal{B}}}(\pmb{x},\pmb{\Omega}_q|\pmb{\Omega}_q^{(i,j-1)})}\\
&&\hspace{-0.9cm}{\rm{s.t.}}\hspace{0.4cm} {\mathcal{U}^{\mathcal{I}}_{n}(\pmb{x},\pmb{\Omega}_{q,n}|\pmb{\Omega}_{q,n}^{(i,j-1)}) \le \bar{C}_n, \hspace{0.2cm}n \in \mathcal{N},}\\
&& \hspace{-0.1cm} {\pmb{\Omega}_{q,n} \succeq 0, \hspace{0.2cm} n \in \mathcal{N}.}
\end{eqnarray}
\end{subequations}
By repeating the procedure (\ref{eq:ortho_opt_quant}) over $j$ until the convergence is attained, the solution $\pmb{\Omega}_q^{(i)}$ is obtained for the $i$th outer loop.
\end{appendices}
\begin{appendices}
\section*{Appendix B: Details of AF Short-term Adaptive Design}  
\renewcommand{\theequation}{B.\arabic{equation}}
\setcounter{equation}{0}  
\renewcommand{\thesection}{B}
\subsection{Optimization over $\pmb{x}$}\label{sec:multi_algoShort_code}
Here, the goal is to optimize the objective function (\ref{eq:multi_optShort}) over the waveform $\pmb{x}^{(i)}$ given the gains $\pmb{p}=\pmb{p}^{(i-1)}$. For this purpose, we apply the MM method. Specifically, at the $j$th iteration of the MM method and the $i$th iteration of the outer loop, the MM method solves a convex QCQP and obtains a solution $\pmb{x}^{(i,j)}$ by substituting the non-convex objective function $\bar{\mathcal{B}}(\pmb{x}, \pmb{p}; \pmb{f})$ with a tight upper bound $\mathcal{U}(\pmb{x},\pmb{p}; \pmb{f}|\pmb{x}^{(i,j-1)})$ around the current iterate $\pmb{x}^{(i,j-1)}$. This bound is obtained by following the same steps as in Appendix \ref{sec:ortho_algo_pro} and is given by
\vspace{-0.2cm}
\begin{eqnarray}\label{eq:multi_Short_code_Bdist} 
&&\hspace{-0.7cm}\mathcal{U}(\pmb{x},\pmb{p}; \pmb{f}|\pmb{x}^{(i,j-1)})\nonumber\\
&&\hspace{-0.7cm}=\phi^{(i,j-1)}\pmb{x}^H\left(\left(\pmb{f}\otimes\pmb{I}_K\right)^H\left(\pmb{P}\otimes\pmb{I}_K\right)\pmb{\Omega}_w\left(\pmb{f}\otimes\pmb{I}_K\right)\right.\nonumber\\
&&\hspace{1cm}\left.+\pmb{\Omega}_z\right)^{-1}\pmb{x}-\mathfrak{Re}\left\{\left(\pmb{d}^{(i,j-1)}\right)^H\pmb{x}\right\},
\end{eqnarray}
where
\begin{eqnarray*}
&&\hspace{-0.7cm}\phi^{(i,j-1)}=\frac{\beta}{1+\beta y^{(i,j-1)}}+\beta(1+0.5\gamma)\nonumber\\
&&\hspace{1.8cm}+\frac{0.5\gamma}{1+\lambda^{(i,j-1)}}\frac{\beta}{\left(1+\beta y^{(i,j-1)}\right)^2};\nonumber\\
&&\hspace{-0.7cm}\pmb{d}^{(i,j-1)}=\left(\frac{2\beta\left(1+0.5\gamma\right)}{1+\beta y^{(i,j-1)}\left(1+0.5\gamma\right)}\right.\nonumber\\
&&\hspace{0.4cm}\left.+2\beta\left(1+0.5\gamma\right)\right)\left(\left(\pmb{f}\otimes\pmb{I}_K\right)^H\left(\pmb{P}\otimes\pmb{I}_K\right)\pmb{\Omega}_w\right.\nonumber\\
&&\hspace{2.4cm}\left.\left(\pmb{f}\otimes\pmb{I}_K\right)+\pmb{\Omega}_z\right)^{-1}\pmb{x}^{(i,j-1)};\nonumber\\
&&\hspace{-0.7cm}\beta=\pmb{f}^H\pmb{P}\pmb{\Sigma}_c\pmb{f};\nonumber\\
&&\hspace{-0.7cm}\gamma=\frac{\pmb{f}^H\pmb{P}\pmb{\Sigma}_t\pmb{f}}{\beta};\nonumber\\
&&\hspace{-0.7cm}y^{(i,j-1)}=\left(\pmb{x}^{(i,j-1)}\right)^H\left(\left(\pmb{f}\otimes\pmb{I}_K\right)^H\left(\pmb{P}\otimes\pmb{I}_K\right)\pmb{\Omega}_w\right.\nonumber\\
&&\hspace{2.4cm}\left.\left(\pmb{f}\otimes\pmb{I}_K\right)+\pmb{\Omega}_z\right)^{-1}\pmb{x}^{(i,j-1)};\nonumber\\
&&\hspace{-0.7cm}\lambda^{(i,j-1)}=\gamma-\frac{\gamma}{1+\beta y^{(i,j-1)}}.\nonumber
\end{eqnarray*}
At the $j$th iteration of the MM method and the $i$th outer loop, we evaluate the new iterate $\pmb{x}^{(i,j)}$ by solving the following QCQP problem
\vspace{-0.2cm} 
\begin{subequations}\label{eq:multi_optShort_code}
\begin{eqnarray}
&&\hspace{-1cm}\pmb{x}^{(i,j)} \leftarrow {\mathop {\text{argmin} }\limits_{\pmb{x}}} \hspace{0.6cm}{\mathcal{U}(\pmb{x},\pmb{p}; \pmb{f}|\pmb{x}^{(i,j-1)})}\\
&& \hspace{0.5cm}{\rm{s.t.}}\hspace{1cm} {\pmb{x}^H\pmb{x} \le P_T.}
\end{eqnarray}
\end{subequations}
The MM method obtains the solution $\pmb{x}^{(i)}$ for the $i$th iteration of the outer loop by solving the problem (\ref{eq:multi_optShort_code}) iteratively over $j$ until a convergence criterion is satisfied. 
\subsection{Optimization over $\pmb{p}$}\label{sec:multi_algoShort_power} 
We consider now the optimization of the gains $\pmb{p}^{(i)}$, when the waveform $\pmb{x}=\pmb{x}^{(i)}$ is given. Similar to the optimization over $\pmb{x}^{(i)}$ in the previous section, we also use the MM method for the optimization over $\pmb{p}$. Towards this goal, we obtain the upper bound $\mathcal{U}(\pmb{x},\pmb{p}; \pmb{f}|\pmb{p}^{(i,j-1)})$ of the objective function $\bar{\mathcal{B}}(\pmb{x}, \pmb{p}; \pmb{f})$ around the current iterate $\pmb{p}^{(i,j-1)}$. This bound is derived by linearizing the difference-of-convex functions via the first-order Taylor approximation \cite{Hunter04Amer}. The bound can then be obtained in (\ref{eq:multi_Short_power_Bdist}) at the top of the next page. 
\begin{figure*}[!t]
\normalsize
\setcounter{equation}{2}
\begin{eqnarray}\label{eq:multi_Short_power_Bdist}
&&\hspace{-1cm}\mathcal{U}(\pmb{x},\pmb{p}; \pmb{f}|\pmb{p}^{(i,j-1)})\nonumber\\
&&\hspace{-1cm}=-\ln\left|\pmb{f}^H\pmb{P}\left(0.5\pmb{\Sigma}_t+\pmb{\Sigma}_c\right)\pmb{f}\pmb{x}\pmb{x}^H+\left(\pmb{f}\otimes\pmb{I}_K\right)^H\left(\pmb{P}\otimes\pmb{I}_K\right)\pmb{\Omega}_w\left(\pmb{f}\otimes\pmb{I}_K\right)+\pmb{\Omega}_z\right|\nonumber\\
&&\hspace{-0.1cm}+0.5\text{tr}\left\{\left(\pmb{f}^H\pmb{P}^{(i,j-1)}\left(\pmb{\Sigma}_t+\pmb{\Sigma}_c\right)\pmb{f}\pmb{x}\pmb{x}^H+\left(\pmb{f}\otimes\pmb{I}_K\right)^H\left(\pmb{P}^{(i,j-1)}\otimes\pmb{I}_K\right)\pmb{\Omega}_w\left(\pmb{f}\otimes\pmb{I}_K\right)+\pmb{\Omega}_z\right)^{-1}\right.\nonumber\\
&&\hspace{4cm}\left.\times\left(\pmb{f}^H\pmb{P}\left(\pmb{\Sigma}_t+\pmb{\Sigma}_c\right)\pmb{f}\pmb{x}\pmb{x}^H+\left(\pmb{f}\otimes\pmb{I}_K\right)^H\left(\pmb{P}\otimes\pmb{I}_K\right)\pmb{\Omega}_w\left(\pmb{f}\otimes\pmb{I}_K\right)\right)\right\}\nonumber\\
&&\hspace{1cm}+0.5\text{tr}\left\{\left(\pmb{f}^H\pmb{P}^{(i,j-1)}\pmb{\Sigma}_c\pmb{f}\pmb{x}\pmb{x}^H+\left(\pmb{f}\otimes\pmb{I}_K\right)^H\left(\pmb{P}^{(i,j-1)}\otimes\pmb{I}_K\right)\pmb{\Omega}_w\left(\pmb{f}\otimes\pmb{I}_K\right)+\pmb{\Omega}_z\right)^{-1}\right.\nonumber\\
&&\hspace{4cm}\left.\times\left(\pmb{f}^H\pmb{P}\pmb{\Sigma}_c\pmb{f}\pmb{x}\pmb{x}^H+\left(\pmb{f}\otimes\pmb{I}_K\right)^H\left(\pmb{P}\otimes\pmb{I}_K\right)\pmb{\Omega}_w\left(\pmb{f}\otimes\pmb{I}_K\right)\right)\right\}.
\end{eqnarray}
\hrulefill
\vspace{-0.3cm}
\end{figure*}
Then, the new iterate $\pmb{p}^{(i,j)}$ at the $j$th iteration of the MM method and the $i$th iteration of the outer loop can be obtained by solving the following optimization problem: 
\vspace{-0.2cm} 
\begin{subequations}\label{eq:multi_optShort_power}
\begin{eqnarray}
&&\hspace{-1cm}\pmb{p}^{(i,j)}\leftarrow{\mathop {\text{argmin} }\limits_{\pmb{p}}} \hspace{0.6cm}{\mathcal{U}(\pmb{x},\pmb{p}; \pmb{f}|\pmb{p}^{(i, j-1)})}\\
&& \hspace{0.5cm}{\rm{s.t.}}\hspace{0.8cm} {\pmb{1}^T\pmb{p} \le P_R,}\\
&& \hspace{1.7cm} {p_n \ge 0, \hspace{0.2cm}n \in \mathcal{N}.}
\end{eqnarray}
\end{subequations}
By repeating the procedure (\ref{eq:multi_optShort_power}) over $j$ until a convergence criterion is satisfied, the solution $\pmb{p}^{(i)}$ is determined for the $i$th outer loop. 
\subsection{Summary of the Proposed Algorithm \ref{al2}}\label{sec:al2}
In summary, in order to solve problem (\ref{eq:multi_optShort}), we propose an algorithm (described in Table Algorithm $\ref{al2}$) that alternates between the optimization over $\pmb{x}$, described in Appendix \ref{sec:multi_algoShort_code} and the optimization over $\pmb{p}$, discussed in Appendix \ref{sec:multi_algoShort_power}. In particular, at the $i$th iteration of the outer loop, the iterate $\pmb{x}^{(i)}$ is obtained by solving a sequence of convex problems (Appendix \ref{sec:multi_algoShort_code}) via the MM method for a fixed $\pmb{p}=\pmb{p}^{(i-1)}$. Then, the iterate $\pmb{p}^{(i)}$ is found by solving a sequence of convex problems (Appendix \ref{sec:multi_algoShort_power}) via the MM method with $\pmb{x}=\pmb{x}^{(i)}$ attained in the previous step. According to the the properties of the MM method \cite{Hunter04Amer, Luo13SIAM}, the proposed scheme yields feasible iterates and a non-increasing objective function along the outer and inner iterations, hence ensuring convergence of the cost function. 
\end{appendices}
\begin{appendices}
\section*{Appendix C: Details of AF Long-term Adaptive Design}  
\renewcommand{\theequation}{C.\arabic{equation}}
\setcounter{equation}{0}  
\renewcommand{\thesection}{C}
\subsection{Optimization over $\pmb{x}$}\label{sec:multi_algoLong_code}
Following the SSUM scheme, at the $j$th inner iteration and the $i$th outer iteration, we optimize the waveform $\pmb{x}^{(i,j)}$ given $\pmb{p}=\pmb{p}^{(i-1)}$ by solving the following convex problem
\begin{subequations}\label{eq:multi_optLong_code}
\begin{eqnarray}
&&\hspace{-1.5cm}\pmb{x}^{(i,j)} \leftarrow {\mathop {\text{argmin} }\limits_{\pmb{x}}} \hspace{0.4cm}{\frac{1}{j}\sum_{l=1}^{j}\mathcal{U}^{(l)}(\pmb{x},\pmb{p}; \pmb{f}^{(l)}|\pmb{x}^{(i,l-1)})}\label{eq:multi_optLong_codeObj}\\
&& \hspace{0cm}{\rm{s.t.}}\hspace{0.65cm} {\pmb{x}^H\pmb{x} \le P_T,}
\end{eqnarray}
\end{subequations}
where $\pmb{f}^{(l)}$ denotes a channel vector $\pmb{f}$ for the fusion center that is randomly and independently generated at the $l$th iteration according to the known distribution of $\pmb{f}$, and $\mathcal{U}^{(l)}(\pmb{x},\pmb{p}; \pmb{f}^{(l)}|\pmb{x}^{(i,l-1)})$ is the locally tight convex upper bound (\ref{eq:multi_Short_code_Bdist}) on the negative Bhattacharyya distance around the point $\pmb{x}^{(i,l-1)}$. Note that the cost function (\ref{eq:multi_optLong_codeObj}) depends on all the realizations of the channel vectors $\pmb{f}^{(l)}$ for $l=1, \dots, j$. The solution $\pmb{x}^{(i)}$ for the $i$th iteration of the outer loop is obtained by solving the problem (\ref{eq:multi_optLong_code}) iteratively over $j$, until a convergence criterion is satisfied. 
\subsection{Optimization over $\pmb{p}$}\label{sec:multi_algoLong_power}
With the optimized waveform $\pmb{x}=\pmb{x}^{(i)}$, SSUM calculates the iterates $\pmb{p}^{(i,j)}$ by solving iteratively the following problems
\begin{subequations}\label{eq:multi_optLong_power}
\begin{eqnarray}
&&\hspace{-1cm}\pmb{p}^{(i,j)}\leftarrow{\mathop {\text{argmin} }\limits_{\pmb{p}}} \hspace{0cm}{\frac{1}{j}\sum_{l=1}^{j}\mathcal{U}^{(l)}(\pmb{x},\pmb{p}; \pmb{f}^{(l)}|\pmb{p}^{(i, l-1)})}\\
&& \hspace{0.5cm}{\rm{s.t.}}\hspace{0.2cm} {\pmb{1}^T\pmb{p} \le P_R,}\\
&& \hspace{1.15cm} {p_n \ge 0, \hspace{0.2cm}n \in \mathcal{N},}
\end{eqnarray}
\end{subequations}
where $\mathcal{U}^{(l)}(\pmb{x},\pmb{p}; \pmb{f}^{(l)}|\pmb{p}^{(i, l-1)})$ is the convex upper bound (\ref{eq:multi_Short_power_Bdist}) on the negative Bhattacharyya distance around the point $\pmb{p}^{(i,l-1)}$. The iterate $\pmb{p}^{(i)}$ is obtained by solving the problem (\ref{eq:multi_optLong_power}) iteratively over $j$ until convergence of the cost function.
\end{appendices}

\bibliographystyle{IEEEtran}
\bibliography{JSAref}

\end{document}